\documentclass[fp,twocolumn]{jpsj3}
\usepackage{txfonts}

\usepackage{graphicx}
\usepackage{bm}
\usepackage{color}
\usepackage{mathrsfs}

\title{Bipolar thermoelectric effects in semiconducting carbon nanotubes:\\
Description in terms of one-dimensional Dirac electrons}

\author{Takahiro Yamamoto$^1$ and Hidetoshi Fukuyama$^2$}
\inst{$^1$ Department of Liberal Arts, Faculty of Engineering, Tokyo University of Science, Katsushika, Tokyo 125-8585, Japan \\
$^2$ Department of Applied Physics, Faculty of Science, Tokyo University of Science, Shinjuku, Tokyo 162-8601, Japan} 

\abst{
The thermoelectric effects in semiconducting single-walled carbon nanotubes (SWCNTs) are investigated based on
the linear response theory combined with the thermal Green's function method. It is shown that the electronic states near 
the lowest conduction band minimum and the highest valence band maximum can be effectively described in terms of 
one-dimensional (1D) Dirac electrons to which a theoretical scheme is developed to describe the thermoelectric responses 
making it possible to study the effects of inter-band impurity scattering and in-gap states. 
Using the proposed scheme, the bipolar thermoelectric effects ({\it i.e.}, the sign inversion of 
the Seebeck coefficient) in semiconducting SWCNTs observed in recent experiments are explained. Moreover, the
temperature dependence of the Seebeck coefficient of semiconducting SWCNTs at low temperature is clarified.
}
 
\kword{Seebeck coefficient, power factor, Dirac electron, carbon nanotube, linear response theory, thermal Green's function}

\begin{document}
\maketitle

\section{Introduction~\label{sec:1}}

The development of high-performance thermoelectric materials is important for sustainable energy production. 
Hicks and Dresselhaus ~\cite{rf:hicks} proposed that 
significant enhancements in the thermoelectric performance of materials could be realized by employing one-dimensional (1D) semiconductors. Various 1D materials exhibiting high thermoelectric 
performance have since been discovered~\cite{rf:lin, rf:rabin, rf:heremans, rf:boukai, rf:hochbaum}. Single-walled carbon nanotubes 
(SWCNTs), which are rolled up graphene in cylindrical form, have received particular interest as high-performance and flexible thermoelectric 1D materials~\cite{rf:small,rf:nakai,rf:hayashi1,rf:hayashi2,rf:Avery,rf:yanagi,rf:hung,rf:Shimizu,rf:nonoguchi1,rf:nonoguchi2,rf:nonoguchi3,rf:fujigaya1,rf:fujigaya2, rf:horike, rf:macLeod,rf:Jiang,rf:ohnishi,rf:TY-HF}. 
At present, the reported maximum power factor of SWCNTs is $\sim$700~$\mu$Wm$^{-1}$K$^{-2}$ at 298 K,~\cite{rf:macLeod} 
which is comparable to that of high-performance inorganic thermoelectric materials. 

Recently, the present authors theoretically demonstrated that ``band-edge engineering" is crucial for the development of high-performance thermoelectric 
materials using impurity-doped semiconducting SWCNTs as an example~\cite{rf:TY-HF}. 
For band-edge engineering, the chemical potential $\mu$ of SWCNTs has been experimentally adjusted via 
chemical adsorption on the SWCNT surface~\cite{rf:nakai,rf:nonoguchi1,rf:fujigaya2}, encapsulation of molecules inside an SWCNT~\cite{rf:fujigaya1}, 
and carrier injection into an SWCNT by applying a gate voltage using a field-effect transistor (FET) setup~\cite{rf:yanagi,rf:Shimizu}. 
In view of the FET experiment~\cite{rf:yanagi,rf:Shimizu}, which clarified the bipolar thermoelectric effect of 
SWCNTs ({\it i.e.}, the sign inversion of the Seebeck coefficient from positive (p-type) to negative (n-type) when the gate voltage is changed),
the present study investigates this bipolar effect theoretically based on the conduction and valence bands, described as
1D Dirac electrons in order to treat the effects of disorder induced coupling between the bands in a unified way~\cite{rf:HFOKS} (see Appendix~\ref{sec:A}).

\section{Theory of Thermoelectric Effects of Dirac Electrons in One-Dimensional Solids~\label{sec:2}}
\subsection{Dirac electrons in one-dimensional solids with disorder~\label{sec:2-1}}
This section gives a brief review of 1D Dirac electrons in solids with disorder.
The Hamiltonian of 1D Dirac electrons in an impurity disorder potential is given by
\begin{eqnarray}
{\mathscr H}=\int_{-\infty}^\infty\!\!\!dx~\Psi^\dagger(x,t) H(x)\Psi(x,t)
\label{eq:hamiltonian}
\end{eqnarray}
with the local Hamiltonian matrix
\begin{eqnarray}
H(x)=-i\hbar v\sigma_x\frac{\partial}{\partial x}+\Delta\sigma_z+U(x),
\label{eq:Hx}
\end{eqnarray}
where $\hbar$ is the Dirac constant, $\Delta$ is one half of the band gap, $v$ is the velocity of a Dirac electron 
in the high-energy region of $|E|\gg\Delta$, $\sigma_x$ and $\sigma_z$ are the $x$ and $z$ components of Pauli matrices, 
\begin{eqnarray}
\sigma_x=\left(
\begin{matrix}
0 & 1\\
1 & 0\\
\end{matrix}
\right)\quad{\rm and}\quad
\sigma_z=\left(
\begin{matrix}
1 & 0\\
0 & -1\\
\end{matrix}
\right),
\end{eqnarray}
respectively, and $U(x)$ is an impurity potential.

In Eq.~(\ref{eq:hamiltonian}),  the field operators $\Psi(x,t)$ and $\Psi^\dagger(x,t)$ are defined as
the following column and row vectors,
\begin{eqnarray}
\Psi(x,t)\equiv \left(
\begin{matrix}
\psi_{1}(x,t) \\
\psi_{2}(x,t) \\
\end{matrix}
\right)
\quad{\rm and}\quad
\Psi^\dagger(x,t)\equiv(\psi_{1}^\dagger(x,t), \psi_{2}^\dagger(x,t)),
\end{eqnarray}
respectively. Here, $\psi_n^\dagger(x,t)$ and $\psi_n(x,t)$ are the fermionic creation and annihilation field operators in the 
Heisenberg picture and satisfy the Heisenberg equations $i\hbar\frac{d\psi_n(x,t)}{dt}=[\psi_n(x,t), {\mathscr H}]$ and 
$i\hbar\frac{d\psi_n^\dagger(x,t)}{dt}=[\psi_n^\dagger(x,t), {\mathscr H}]$, respectively. Thus, it can be easily proven that 
$\Psi(x,t)$ and $\Psi^\dagger(x,t)$ satisfy the Shr{\" o}dinger equation
\begin{eqnarray}
i\hbar\frac{d\Psi(x,t)}{dt}=H(x)\Psi(x,t)
\label{eq:sch_eq1}
\end{eqnarray}
and its Hermitian conjugate
\begin{eqnarray}
-i\hbar\frac{d\Psi^\dagger(x,t)}{dt}=\left(H(x)\Psi(x,t)\right)^\dagger.
\label{eq:sch_eq2}
\end{eqnarray}

Performing the Fourier transform of $\Psi^\dagger(x,t)$ as
\begin{eqnarray}
\Psi^\dagger(x,t)=\frac{1}{\sqrt{L}}\sum_k e^{-ikx}\Phi_{k}^\dagger(t)
\label{eq:psi_f}
\end{eqnarray}
with $\Phi_{k}^\dagger(t)\equiv(c_{1k}^\dagger(t), c_{2k}^\dagger(t))$ and using 
\begin{eqnarray}
U(x)=\sum_q e^{iqx} U(q),
\label{eq:uxuk}
\end{eqnarray}
the 1D Dirac Hamiltonian in Eq.~(\ref{eq:hamiltonian}) can be rewritten as
\begin{eqnarray}
{\mathscr H}=\sum_k \Phi^\dagger_k H_0(k)\Phi_k+\sum_{k,q} \Phi^\dagger_{k+q} U(q)\Phi_k,
\label{eq:H_k}
\end{eqnarray}
where $H_0(k)$ is the Hamiltonian density of a 1D free Dirac electron in $k$ space, which is given by 
\begin{eqnarray}
H_0(k)&=&\hbar vk\sigma_x+\Delta\sigma_z\\
&=&\left(
\begin{matrix}
\Delta & \hbar vk \\
\hbar vk & -\Delta \\
\end{matrix}
\right).
\label{eq:1d_dirac_ham_k}
\end{eqnarray}
The eigenvalues of $H_0(k)$ in Eq.~(\ref{eq:1d_dirac_ham_k}) can be easily obtained as
\begin{eqnarray}
E_{\pm}(k)=\pm\sqrt{\Delta^2+(\hbar vk)^2},
\end{eqnarray}
where $\pm$ corresponds the conduction ($+$) and valence ($-$) bands, respectively. Thus, the band gap
is given by $E_{\rm g}\equiv E_{+}(0)-E_{-}(0)=2\Delta$.

\subsection{General theory of thermoelectric responses~\label{sec:2-2}}
The thermoelectric effect is typically characterized by the Seebeck coefficient, $S$, which is defined as the voltage induced by 
a finite temperature gradient along a given direction (herein the $x$-direction) under the condition that there is no electrical current 
({\it i.e.}, $J=0$) along that direction. This can be written as
\begin{eqnarray}
S\equiv-\left(\frac{\Delta V}{\Delta T}\right)_{J=0},
\label{eq:S_def}
\end{eqnarray}
where $\Delta V$ is the induced voltage and $\Delta T$ is the temperature difference between the two ends of the material.

In the presence of both an electric field $\mathcal{E}$ and a temperature gradient $dT/dx$ along the $x$-direction, the current density 
$J$ is generally given by
\begin{eqnarray}
J=L_{11}\mathcal{E}-\frac{L_{12}}{T}\frac{dT}{dx}
\label{eq:J}
\end{eqnarray}
within the linear response regime with respect to $\mathcal{E}$ and $dT/dx$. Here, $L_{11}$ and $L_{12}$ are 
the electrical conductivity and the thermoelectrical conductivity, respectively.
The zero-current condition ($J=0$) leads to 
$L_{11}\mathcal{E}=\frac{L_{12}}{T}\frac{dT}{dx}$. Because the electric field and the temperature gradient can be written as 
$\mathcal{E}=-\Delta V/L$ and $dT/dx=\Delta T/L$, respectively, for a spatially uniform system with length $L$ (assumed here), 
$S$ as defined in Eq.~(\ref{eq:S_def}) can be expressed in terms of the response functions $L_{11}$ and $L_{12}$ as
\begin{eqnarray}
S=\frac{1}{T}\frac{L_{12}}{L_{11}}.
\label{eq:S}
\end{eqnarray}
One of the figures of merit for thermoelectric materials is the power factor, $PF$, defined as
\begin{eqnarray}
PF\equiv\sigma S^2=\frac{1}{T^2}\frac{L_{12}^2}{L_{11}}.
\label{eq:PF}
\end{eqnarray}
It should be noted that the basic quantities used for the thermoelectric responses of materials are $L_{11}$ and $L_{12}$ rather than $S$ and $PF$.

\subsection{Microscopic representation of $L_{11}$ and $L_{12}$~\label{sec:2-3}}
$L_{11}$ and $L_{12}$ are expressed in terms of the $J_{\rm e}$-$J_{\rm e}$ correlation function and 
the $J_{\rm e}$-$J_{\rm Q}$ correlation function, respectively, where $J_{\rm e}$ and $J_{\rm Q}$ are the electronic current and 
the thermal current, respectively~\cite{rf:langer,rf:ambegaokar}. Microscopic justification for calculating $L_{11}$ based on the $J_{\rm e}$-$J_{\rm e}$ correlation function
and $L_{12}$ based on the $J_{\rm e}$-$J_{\rm Q}$ correlation function was first given by Kubo in 1957~\cite{rf:kubo} and Luttinger in 1964~\cite{rf:Luttinger}, respectively.
This subsection presents the explicit expressions of $L_{11}$ and $L_{12}$ for 1D Dirac electrons with an impurity potential 
based on the thermal Green's function formalism.

According to the linear response theory, the thermoelectrical conductivity $L_{12}$ can be obtained as
\begin{eqnarray}
& &L_{12}=-\lim_{\omega\to{0}}\frac{\chi^{\rm R}_{12}(\omega)-\chi^{\rm R}_{12}(0)}{i\omega},\\
& &\chi_{12}^{\rm R}(\omega)=\chi_{12}(i\omega_\lambda)\Big|_{i\omega_\lambda\to\hbar\omega+i\delta},
\end{eqnarray}
where $\chi_{12}(i\omega_\lambda)$ is the correlation function between the electrical current $J_{\rm e}$ and 
the thermal current $J_{\rm Q}$, expressed as
\begin{eqnarray}
\chi_{12}(i\omega_\lambda)=\frac{1}{V}\int_0^\beta\!\!\! d\tau 
\left\langle
T_{\tau} \{J_{\rm e}(\tau)J_{\rm Q}(0)\}\right\rangle
e^{i\omega_\lambda\tau},
\end{eqnarray}
where $\beta\equiv 1/(k_{\rm B}T)$ is the inverse temperature, $T_{\tau}$ is the imaginary-time-ordering operator, 
$\langle\cdots\rangle$ denotes the thermal average, and $V$ is the volume of a system.

Now, $J_{\rm e}(\tau)$ and $J_{\rm Q}(\tau)$ for 1D Dirac electrons
with a disorder potential are needed. The electric current carried by 1D Dirac electrons is given by 
\begin{eqnarray}
J_{\rm e}(t)=-e\sum_{k}\Phi_k^\dagger(t)v(k)\Phi_k(t),
\label{eq:j_J_e_x}
\end{eqnarray} 
where $e(>0)$ is the elementary charge and $v(k)$ is a $2\times 2$ velocity matrix, which is given by
\begin{eqnarray}
v(k)=\frac{1}{\hbar}\frac{\partial H_0(k)}{\partial k}
=v\sigma_x
=\left(
\begin{matrix}
0 & v \\
v & 0 \\
\end{matrix}
\right).
\label{eq:current_mat}
\end{eqnarray}
The thermal current $J_{\rm Q}(t)$ is expressed as $J_{\rm Q}(t)=J_{\rm E}(t)+\frac{\mu}{e}J_{\rm e}(t)$, where
the energy current $J_{\rm E}(t)$ is defined as 
\begin{eqnarray}
J_{\rm E}(t)=\frac{d{\mathscr A}(t)}{dt}
\label{eq:dAdt}
\end{eqnarray}
with energy polarization
\begin{eqnarray}
{\mathscr A}(t)=\frac{1}{2}\int_{-\infty}^\infty\!\!\!dx~\Psi^\dagger(x,t)\left(\frac{\sin Qx}{Q} H(x)+H(x)\frac{\sin Qx}{Q}\right) \Psi(x,t).
\label{eq:A}
\end{eqnarray}
Here, $Q$ is a parameter used to control the effects of the unbounded variable $x$ and should be set to zero in the final step of a calculation~\cite{rf:ogata-fukuyama2018}.
Using Eqs.~(\ref{eq:sch_eq1}) and (\ref{eq:sch_eq2}), Eq.~(\ref{eq:dAdt}) is calculated as
\begin{eqnarray}
J_{\rm E}(t)=\frac{v}{2}\int_{-\infty}^\infty\!\!\!dx~\Psi^\dagger(x,t)\left\{
\sigma_xH(x)+H(x)\sigma_x
\right\}\Psi(x,t).
\label{eq:j_Q_x}
\end{eqnarray} 
The derivation of Eq.~(\ref{eq:j_Q_x}) is given in Appendix~\ref{sec:B}.
Substituting Eqs.~(\ref{eq:psi_f}) and (\ref{eq:uxuk}) into Eq.~(\ref{eq:j_Q_x}) yields
\begin{eqnarray}
& &J_{\rm E}(t)=\frac{1}{2}\sum_{k}\left[\Phi^\dagger_{k}(t) \left\{v(k)H_0(k)+H_0(k)v(k)\right\} \Phi_{k}(t)\right.\nonumber\\
& &+\sum_{q}\left.\Phi^\dagger_{k+q}(t) \left\{v(k+q)U(q)+ U(q)v(k)\right\} \Phi_{k}(t)\right],
\label{eq:j_Q}
\end{eqnarray}
where $H_0(k)$ is the Hamiltonian density of 1D free Dirac electrons in Eq.~(\ref{eq:1d_dirac_ham_k}).

In the imaginary-time Heisenberg picture ($t\to -i\tau$), the electric current is expressed as
\begin{eqnarray}
J_{\rm e}(\tau)=-e\int_{-\infty}^\infty\!\!\!dx~{\bar \Phi}_k(\tau)v(k)\Phi_k(\tau),
\label{eq:j_J_e}
\end{eqnarray} 
where ${\bar \Phi}_{k}(\tau)\equiv({\bar c}_{1k}(\tau), {\bar c}_{2k}(\tau))=(e^{\tau{H}}c_{1k}^\dagger{e}^{-\tau{H}}, e^{\tau{H}}c_{2k}^\dagger{e}^{-\tau{H}})$. 
Similarly, the energy current is given by
\begin{eqnarray}
J_{\rm E}(\tau)&=&-\frac{1}{2e}\sum_{k}\left[\left\{{\bar \Phi}_k(\tau)H_0(k)+\sum_{q}{\bar \Phi}_{k+q}(\tau)U(q)\right\}J_{\rm e}\Phi_k(\tau)\right.\nonumber\\
& &+\left.{\bar \Phi}_k(\tau)J_{\rm e}\left\{H_0(k)\Phi_k(\tau)+\sum_qU(q)\Phi_{k-q}(\tau)\right\}\right],
\label{eq:j_Q_Heisenberg}
\end{eqnarray}
where the current matrix $J_{\rm e}=-ev\sigma_x$. As first noted by Johnson and Mahan~\cite{rf:J-M}, Eq.~(\ref{eq:j_Q_Heisenberg}) can be rewritten as
\begin{eqnarray}
J_{\rm E}(\tau)=-\frac{1}{2e}\sum_{k}\left[
\frac{d{\bar \Phi}_k(\tau)}{d\tau}J_{\rm e}\Phi_k(\tau)-{\bar \Phi_k}(\tau) J_{\rm e}\frac{d\Phi_k(\tau)}{d\tau}
\right]
\label{eq:J-M}
\end{eqnarray}
by using the relations
\begin{eqnarray}
\frac{d\Phi_{k}(\tau)}{d\tau}&=&-H_0(k)\Phi_{k}(\tau)-\sum_{q}U(q)\Phi_{ k-q}(\tau),\\
\frac{d{\bar \Phi}_{nk}(\tau)}{d\tau}&=&{\bar \Phi}_{k}(\tau)H_0(k)+\sum_{q}{\bar \Phi}_{k+q}(\tau) U(q),
\label{eq:dcdt}
\end{eqnarray}
which are derived from $dc_{nk}(\tau)/d\tau=[H,c_{nk}(\tau)]$ and $d{\bar c}_{nk}(\tau)/d\tau=[H,{\bar c}_{nk}(\tau)]$. 
Using Eq.~(\ref{eq:J-M}), the correlation function $\chi_{12}(i\omega_\lambda)$ is 
expressed in terms of the impurity-averaged thermal Green's function ${\mathscr G}(k,i\epsilon_n)\equiv\langle{\mathscr G}(k,k',i\epsilon_n)\rangle_{\rm imp}$ as
\begin{eqnarray}
\chi_{12}(i\omega_\lambda)&=&-\frac{ev^2}{V\beta}\sum_{n}\left(\frac{i\epsilon_n+i\epsilon_{n+}}{2}-\mu\right)\nonumber\\
& &\times\sum_{k}{\rm Tr}\left[
\sigma_x{\mathscr G}(k,i\epsilon_{n}){\tilde\sigma}_x(k){\mathscr G}(k,i\epsilon_{n+})
\right]
\label{eq:chi_12}
\end{eqnarray}
with $\epsilon_{n+}\equiv\epsilon_n+\omega_\lambda$ and ${\tilde \sigma}_x(k)$ is given as ${\tilde \sigma}_x(k)=\sigma_x+\sigma'_x(k)$ with
\begin{eqnarray}
\sigma'_x(k)&=&\sum_{k',k'',k'''}\left\langle
U(k-k'){\mathscr G}(k',k'',i\epsilon_{n}){\tilde \sigma}_x(k'')\right.\nonumber\\
& &\qquad\left.\times~{\mathscr G}(k'',k''',i\epsilon_{n})U(k'''-k)
\right\rangle_{\rm imp}
\label{eq:sigma'x}
\end{eqnarray}
that gives the vertex correction for the current operator. In the following discussion, 
we assume that the vertex correction is neglected, {\it i.e.}, ${\tilde \sigma}_x\to \sigma_x$.
In Eq.~(\ref{eq:chi_12}), ${\mathscr G}(k,i\epsilon_n)$ is given by a 2$\times$2 matrix as
\begin{eqnarray}
{\mathscr G}(k,i\epsilon_n)=\left(
\begin{matrix}
{\mathscr G}_{11}(k,i\epsilon_n) & {\mathscr G}_{12}(k,i\epsilon_n) \\
{\mathscr G}_{21}(k,i\epsilon_n) & {\mathscr G}_{22}(k,i\epsilon_n) \\
\end{matrix}
\right),
\label{eq:thermal_green}
\end{eqnarray}
where ${\mathscr G}(k,i\epsilon_n)$ is defined by the Fourier transform of ${\mathscr G}_{nm}(k,\tau)\equiv -\left\langle{T_\tau \{c_{nk}(\tau){\bar c}_{mk}(0)}\}\right\rangle$,
\begin{eqnarray}
{\mathscr G}_{nm}(k,i\epsilon_n)=\int_0^\beta\!\!\!d\tau~{\mathscr G}_{nm}(k,\tau)e^{i\epsilon_n\tau}.
\label{eq:thermal_green_def}
\end{eqnarray}

The summation over $n$ in Eq.~(\ref{eq:chi_12}) can be transformed into a contour integral in the complex energy space. Taking the limit of $\omega\to{0}$ after the analytic continuation $i\omega_\lambda\to\hbar\omega+i\delta$ yields
the expression of $L_{12}$ as
\begin{eqnarray}
L_{12}=-\frac{1}{e}\int_{-\infty}^{\infty}\!\!\! dE\left(-\frac{\partial f(E-\mu)}{\partial E}\right)(E-\mu)\alpha(E),
\label{eq:Sommerfeld-Bethe}
\end{eqnarray}
where $f(E)=[1+\exp\{(E-\mu)/k_{\rm B}T\}]^{-1}$ is the Fermi-Dirac distribution function and
$\alpha(E)$ is often called the spectral conductivity, which is expressed as
\begin{eqnarray}
\alpha(E)&=&\frac{\hbar e^2v^2}{2\pi V}\sum_{k}{\rm Tr}\left[
\sigma_xG^{\rm A}(k,E){\sigma}_xG^{\rm R}(k,E)\right.\nonumber\\
& &\left.-{\rm Re}\left\{\sigma_xG^{\rm R}(k,E){\sigma}_xG^{\rm R}(k,E)\right\}
\right]
\label{eq:spectral_conductivity}
\end{eqnarray}
by use of the retarded/advanced Green's function,
\begin{eqnarray}
G^{\rm R/A}(E,k)=\left\{
EI-H_0(k)-\Sigma^{\rm R/A}(E,k)
\right\}^{-1},
\end{eqnarray}
where $I$ is the 2$\times$2 identity matrix. The expression of $L_{12}$ in Eq.~(\ref{eq:Sommerfeld-Bethe})
was first proposed for general cases by Sommerfeld and Bethe~\cite{rf:SB} in 1933, and subsequently by 
Mott and Jones~\cite{r:mott-jones} and by Wilson~\cite{rf:wilson} based on Boltzmann transport theory. 
Recently, Ogata and Fukuyama clarified the range of validity of the Sommerfeld and Bethe relation expressed as Eq.~(\ref{eq:Sommerfeld-Bethe}) 
for single-band systems with a disorder potential, electron-phonon coupling, and electron correlations as well as for 
multi-band disorder systems based on the L{\" u}ttinger-Kohn representation~\cite{rf:ogata-fukuyama2018}.

Using the expression of $\alpha(E)$ in Eq.~(\ref{eq:spectral_conductivity}), the electrical conductivity $L_{11}$ of 1D Dirac electrons in a disorder potential 
is given by
\begin{eqnarray}
L_{11}=\int_{-\infty}^{\infty}\!\!\! dE\left(-\frac{\partial f(E-\mu)}{\partial E}\right)\alpha(E).
\label{eq:L_11}
\end{eqnarray}

\section{Electronic States and Thermoelectric Responses of 1D Dirac Electrons}
In this section, the thermoelectric effects of 1D Dirac electrons in a disorder potential are studied based on the thermal Green's function formalism 
using the self-energy corrections of Green's functions. The following subsections present two methods for treating self-energy 
correction: the constant-$\tau$ approximation and the self-consistent Born approximation.

\subsection{Constant-$\tau$ approximation}
As the simplest treatment of self-energy correction for a disorder potential, the constant-$\tau$ approximation ({\it i.e.}, 
$\Sigma^{{\rm R}/{\rm A}}=\mp{i}(\hbar/2\tau)I$) is employed here and vertex correction is neglected. 
Here, $\tau$ is the relaxation time of a Dirac electron scattered by a disorder potential, which is assumed to be independent of the energy $E$ and the wavenumber $k$. 
Although this constant-$\tau$ approximation is simple, the approximation is very useful for obtaining an overview of the thermoelectric response of 
1D Dirac electrons in a disorder potential.

\begin{figure}[t]
  \begin{center}
  \includegraphics[keepaspectratio=true,width=80mm]{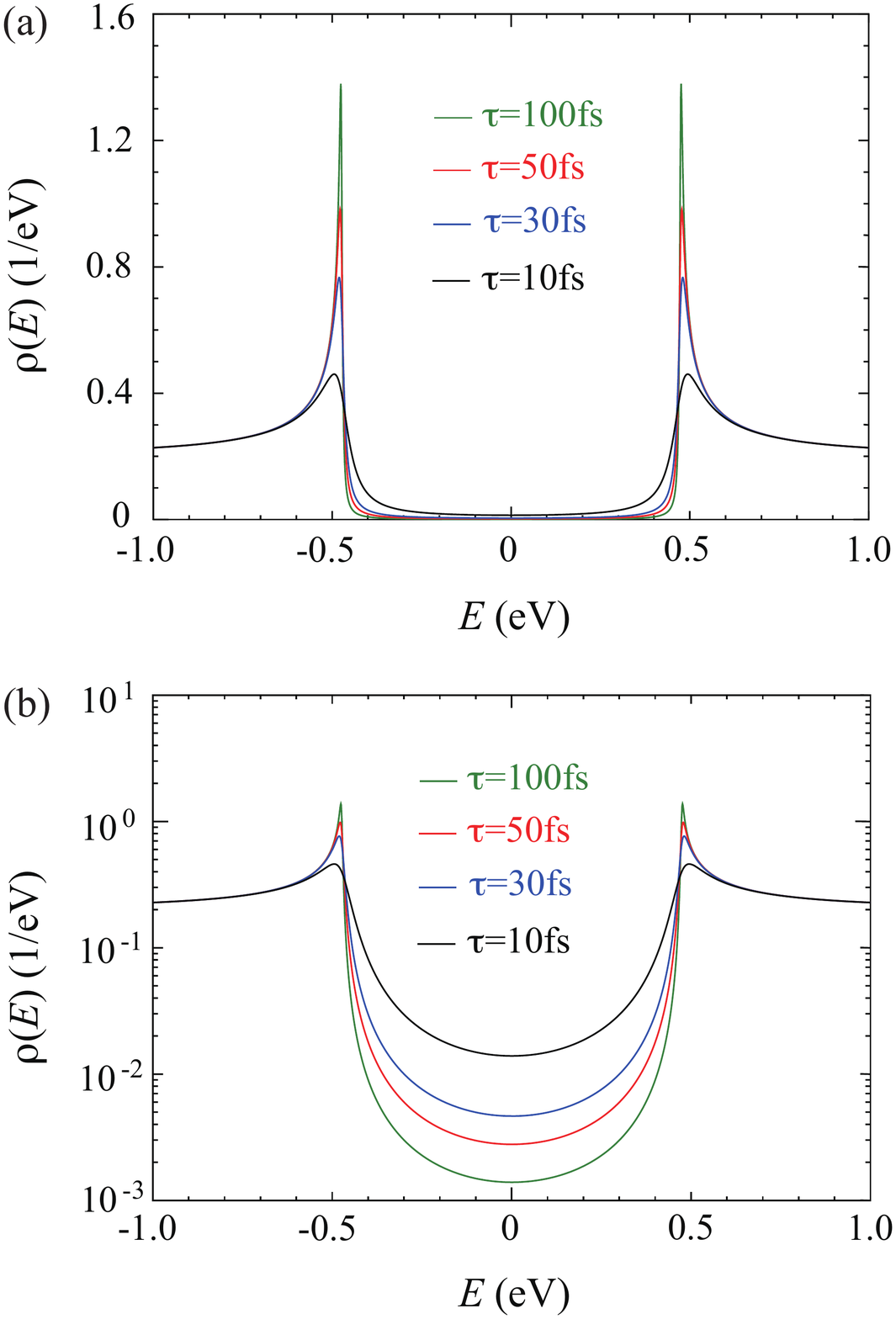}
  \end{center}
\caption{(Color online) (a) Density of states near the band gap of (10,0) SWCNTs 
per spin and orbital for $\tau=10$ (black curve), $30$ (blue curve), $50$ (red curve), and $100$~fs (green curve) calculated using the constant-$\tau$ 
approximation and (b) the corresponding semi-log plots. }
\label{fig:01}
\end{figure}

\subsubsection{Density of states and spectral conductivity}
In the constant-$\tau$ approximation, the retarded and advanced Green's functions 
for a 1D Dirac electron are given by
\begin{eqnarray}
G^{{\rm R}/{\rm A}}(E,k)=\left\{
\left(E\pm i\frac{\hbar}{2\tau}\right)I-(\hbar vk\sigma_x+\Delta\sigma_z)
\right\}^{-1},
\label{eq:const-tau_green}
\end{eqnarray}
which leads to the following expressions for the density of states (DOS) $\rho(E)$ per unit cell of the system and 
spectral conductivity $\alpha(E)$
\begin{eqnarray}
\rho(E)&=&-\frac{1}{\pi N}{\rm Tr}\sum_{k}{\rm Im}~G^{{\rm R}}(k,E)\\
&=&\frac{a}{\pi(\hbar v)^2}{\rm Im}\left(\frac{\gamma+iE}{k_+}\right)
\label{eq:dos_const-tau}
\end{eqnarray}
and
\begin{eqnarray}
\alpha(E)=i\frac{e^2}{\hbar}\frac{1}{\pi A}\frac{1}{k_+-k_-}\left\{
1-\frac{1}{k_+k_-}\frac{E^2+\gamma^2-\Delta^2}{(\hbar v)^2}
\right\},
\label{eq:const-tau_alpha}
\end{eqnarray}
with $\gamma\equiv\hbar/2\tau$. Here, $N$ is the total number of unit cells, $a$ is the length of a unit cell, 
$A$ is the cross-sectional area of the system,
and
\begin{eqnarray}
k_{\pm}^2\equiv\frac{(E\pm i\gamma)^2-\Delta^2}{(\hbar v)^2},\quad {\rm Im}~k_{\pm}>0.
\end{eqnarray}
For an SWCNT, $A$ is conventionally taken to be $A\equiv\pi d_{\rm t}\delta$,  
where $\delta=0.34$ nm is the van der Waals diameter of carbon. 

\begin{figure}[t]
  \begin{center}
  \includegraphics[keepaspectratio=true,width=80mm]{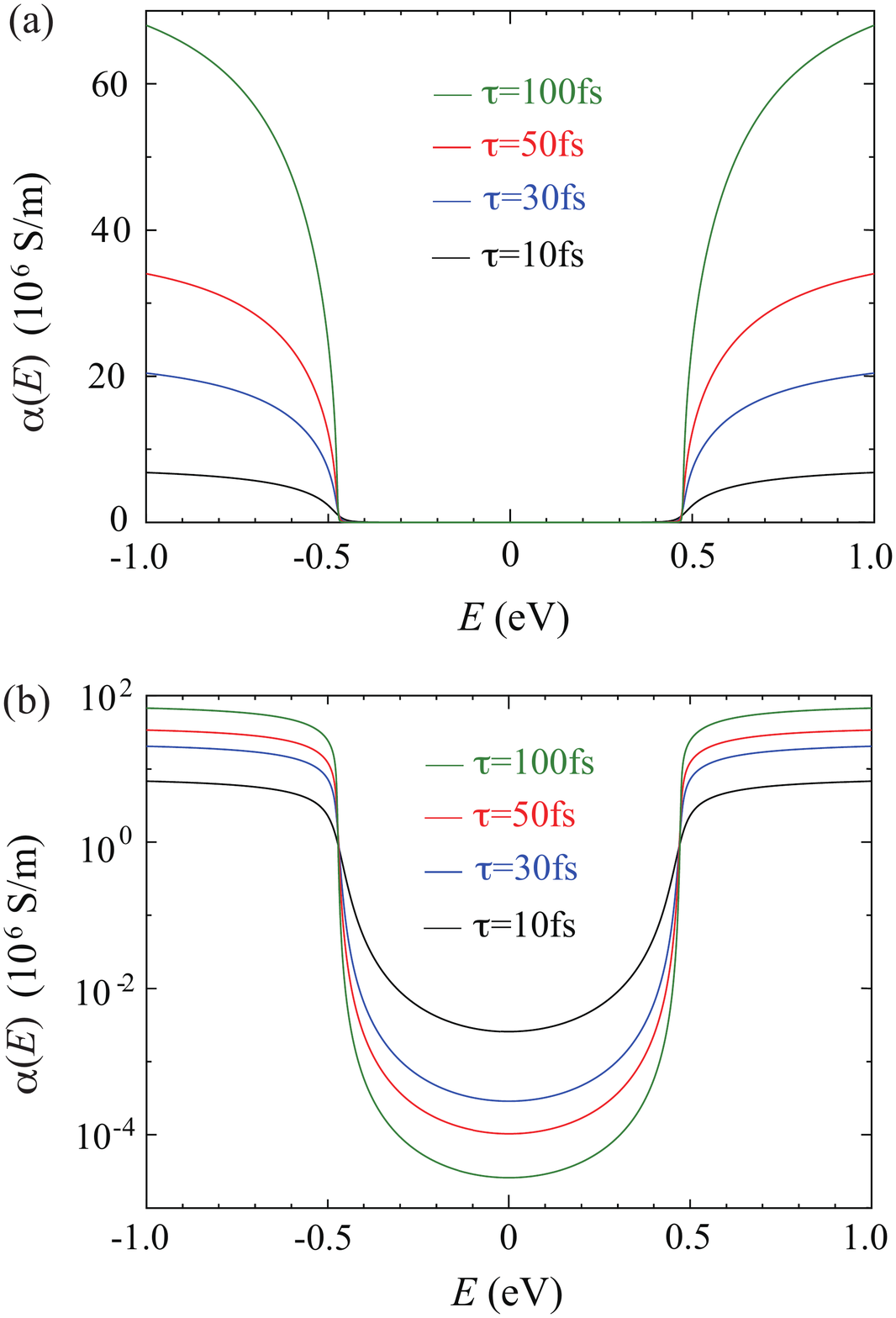}
  \end{center}
\caption{(Color online) (a) Spectral conductivities of (10,0) SWCNTs for $\tau=10$ (black curve), $30$ (blue curve), $50$ (red curve), 
and $100$~fs (green curve) calculated using the constant-$\tau$ approximation and (b) the corresponding semi-log plots. }
\label{fig:02}
\end{figure}

Figures~\ref{fig:01}(a) and \ref{fig:01}(b) show $\rho(E)$ near the band gap of (10,0) SWCNTs 
per spin and orbital for $\tau=10$, $30$, $50$, and $100$~fs ($\gamma$=32.91, 10.97, 6.58, and 3.29~meV) calculated using Eq.~(\ref{eq:dos_const-tau}) and the corresponding semi-log plots, respectively. 
Two sharp peaks of $\rho(E)$ appear around $E=\pm\Delta=\pm 0.475$~eV, corresponding to the van Hove singularity points of 
pristine (10,0) SWCNTs without disorder. The maximum value of $\rho(E)$ decreases with decreasing $\tau$ (increasing $\gamma$)
due to disorder scattering. In the high-energy region of $|E|\gg\Delta$, the $\rho(E)$ data in Figs.~\ref{fig:01}(a) and \ref{fig:01}(b) converge to
the constant value of $\rho(\infty)=a/\pi\hbar v=0.206$/eV, which is independent of $\tau$, where the unit cell length $a$ of a (10,0) SWCNT is 
$a=0.426$~nm and the velocity $v$ of a Dirac electron in the (10,0) SWCNT is given by $v=1.027\times 10^{6}$~m/s. 
In addition, a (10,0) SWCNT with finite $\tau$ exhibits a finite DOS even in its band gap ($|E|<\Delta$), which increases with decreasing $\tau$ 
(increasing $\gamma$), as shown in Fig.~\ref{fig:01}(b). As shown below, the in-gap states have a crucial consequence in the thermoelectric effects of SWCNTs.

Figures~\ref{fig:02}(a) and \ref{fig:02}(b) show the spectral conductivities of (10,0) SWCNTs for $\tau=10$, $30$, $50$, and $100$~fs 
calculated using Eq.~(\ref{eq:const-tau_alpha}) and the corresponding semi-log plots, respectively. Here, $\alpha(E)$ in Eq.~(\ref{eq:const-tau_alpha}) was 
multiplied by a factor of 4 for the (10,0) SWCNTs because their lowest-conduction (LC) and highest-valence (HV) bands both have two-fold orbital degeneracy 
and two-fold spin degeneracy (see Appendix~\ref{sec:A}).
It can be seen that $\alpha(E)$ in $|E|>\Delta$ decreases with decreasing $\tau$, whereas that in $|E|<\Delta$ increases
with decreasing $\tau$ (Fig.~\ref{fig:02}(b)) because the DOS of the in-gap states increases with increasing $\tau$, as shown in Fig.~\ref{fig:01}(b).
Once $\alpha(E)$ is obtained, the  electrical conductivity $L_{11}$ and thermoelectrical conductivity $L_{12}$ 
can be respectively calculated using Eqs.~$(\ref{eq:L_11})$ and (\ref{eq:Sommerfeld-Bethe}), 
and then the Seebeck coefficient $S$ and the power factor $PF$ can be respectively obtained using Eqs.~(\ref{eq:S}) and (\ref{eq:PF}).

\begin{figure}[t]
  \begin{center}
  \includegraphics[keepaspectratio=true,width=80mm]{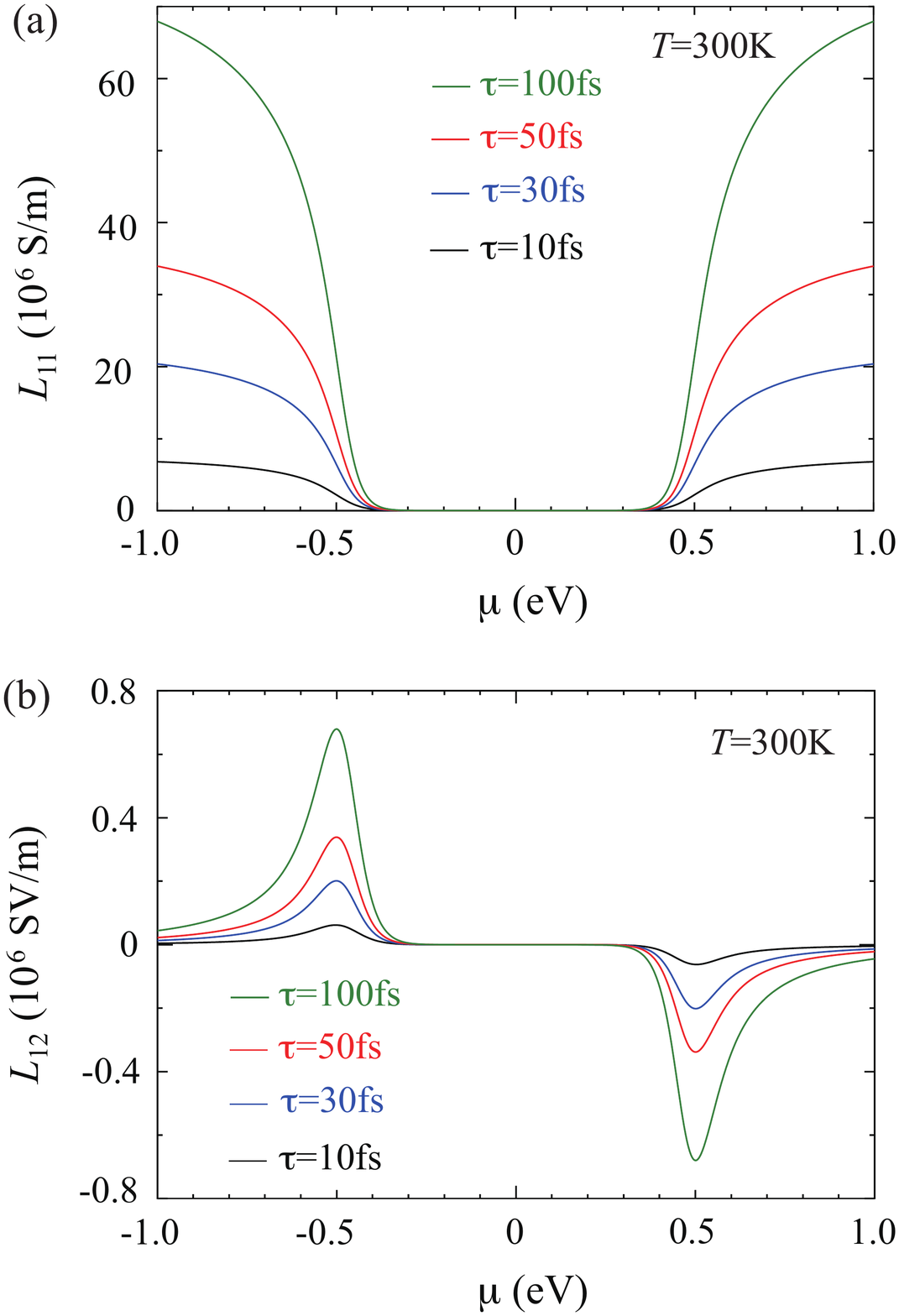}
  \end{center}
\caption{(Color online) Chemical potential dependence of (a) electrical conductivity $L_{11}$ and (b) thermoelectric conductivity $L_{12}$ of (10,0) SWCNTs for 
$\tau=10$ (black curve), $30$ (blue curve), $50$ (red curve), and $100$~fs (green curve) calculated using the constant-$\tau$ approximation. }
\label{fig:03}
\end{figure}

\subsubsection{Chemical potential dependence of $L_{11}$ and $L_{12}$ at 300~K}
Inspired by recent experiments regarding the bipolar thermoelectric effects of SWCNTs using the FET setup~\cite{rf:yanagi},
we study the $\mu$ dependence of $L_{11}$ and $L_{12}$ of SWCNTs at $T=300$~K.
Figure~\ref{fig:03}(a) shows the $\mu$ dependence of the $L_{11}$ of (10,0) SWCNTs 
for $\tau=10$, $30$, $50$, and $100$~fs at $T=300$~K. As expected from Eq.~(\ref{eq:L_11}), 
the $\mu$ dependence of $L_{11}$ for each $\tau$ value shows features similar to those of the spectral conductivity $\alpha(E)$ in Fig.~\ref{fig:02}(a). 
As shown in Fig.~\ref{fig:03}(b), the $L_{12}$ values of (10,0) SWCNTs for $\tau=10$, $30$, $50$, and $100$~fs at $T=300$~K 
have a sharp dip or peak near the conduction and valence band edges ($E=\pm\Delta$), respectively.

\begin{figure}[t]
  \begin{center}
  \includegraphics[keepaspectratio=true,width=80mm]{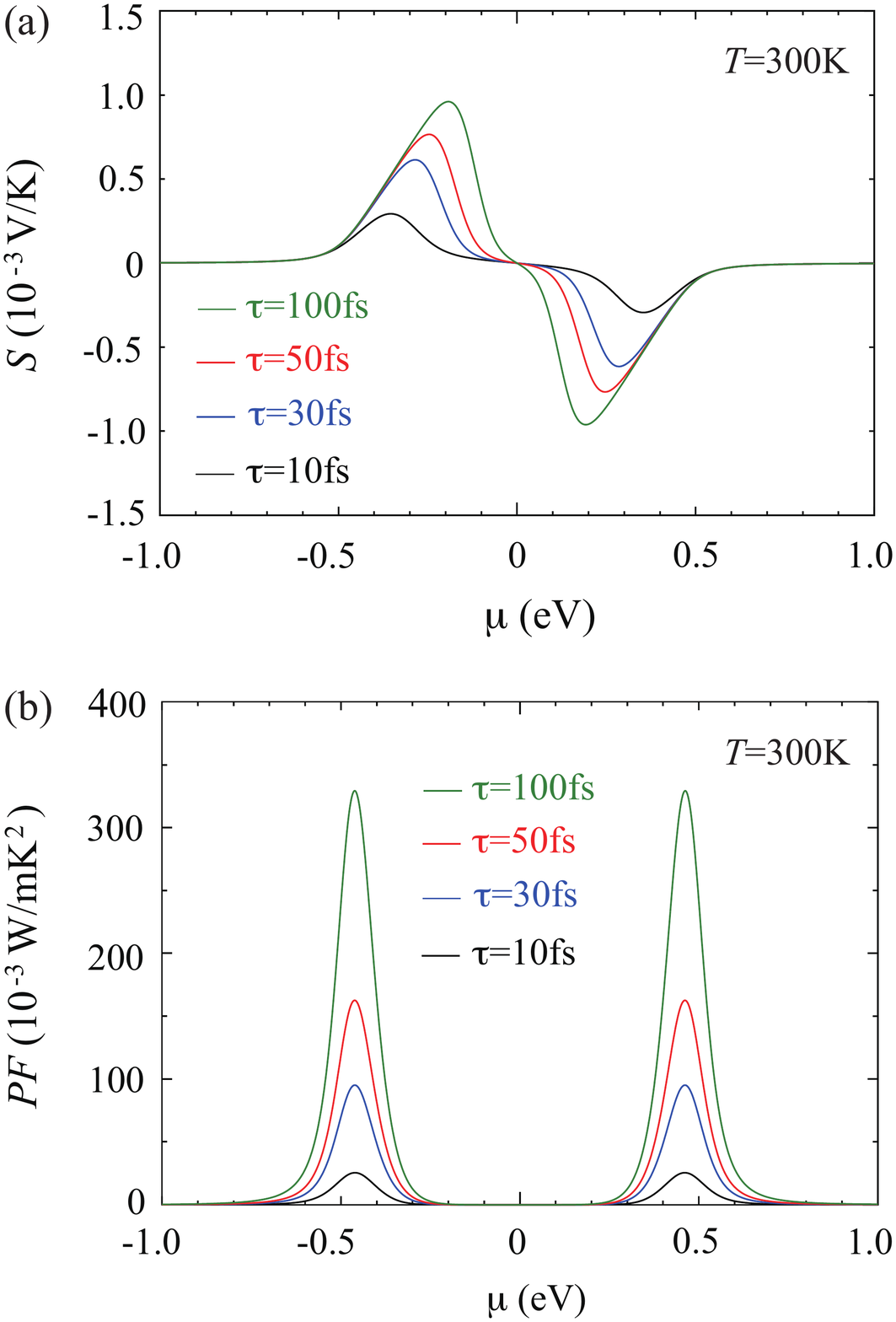}
  \end{center}
\caption{(Color online) Chemical potential dependence of (a) Seebeck coefficient $S$ and (b) power factor $PF$ of (10,0) SWCNTs for 
$\tau=10$ (black curve), $30$ (blue curve), $50$ (red curve), and $100$~fs (green curve) calculated using the constant-$\tau$ approximation. }
\label{fig:04}
\end{figure}

\subsubsection{Chemical potential dependence of $S$ and $PF$ at 300~K}
By substituting the $L_{11}$ and $L_{12}$ data in Fig.~\ref{fig:03} into Eq.~(\ref{eq:S}), the $\mu$ dependence of the Seebeck coefficient $S$ 
of (10,0) SWCNTs can be obtained. As shown in Fig.~\ref{fig:04}(a), the absolute value of $S$ decreases with decreasing $\tau$.
In addition, $S$ exhibits a sign inversion from positive (p-type) to negative (n-type) when $\mu$ changes from 
negative to positive. Such a bipolar thermoelectric effect of SWCNTs was recently observed in experiments using an electric double layer 
transistor~\cite{rf:yanagi,rf:Shimizu}. Moreover, Fig.~\ref{fig:04}(a) shows that $S$ has maximum and minimum values at the optimal
chemical potentials $\mu=\pm\mu_{\rm opt}$, and that $\mu_{\rm opt}$ shifts toward the energy band edge with decreasing $\tau$. 
This trend is different from that reported in a previous study, which found that $\mu_{\rm opt}$ is independent of $\tau$ using the Boltzmann transport theory 
combined with a model of two independent bands (TIBs) under the constant-$\tau$ approximation~\cite{rf:hung}. The $\tau$-dependent shift of $\mu_{\rm opt}$
results from the in-gap states (see Fig.~\ref{fig:01}(b)), which are not taken into account in the framework of the TIB model.

The power factor $PF$ of (10,0) SWCNTs can also be calculated by substituting the $L_{11}$ and $L_{12}$ data in Fig.~\ref{fig:03} into Eq.~(\ref{eq:PF}).
Figure~\ref{fig:04}(b) shows the $\mu$ dependence of the $PF$ of (10,0) SWCNTs for $\tau=10$, $30$, $50$, and $100$~fs at $T=300$~K.
$PF$ has its maximum value near the conduction and valence band edges $E=\pm\Delta(=\pm0.475~{\rm eV})$ of a pristine (10,0) SWCNT, 
and decreases with decreasing $\tau$. The high $PF$, on the order of 100~mW/mK$^2$, at the energy band edges is consistent
with our previous theoretical work on the $\mu$ dependence of the $PF$ of impurity-doped SWCNTs using a single-band approximation~\cite{rf:TY-HF}. 

\begin{figure}[t]
  \begin{center}
  \includegraphics[keepaspectratio=true,width=80mm]{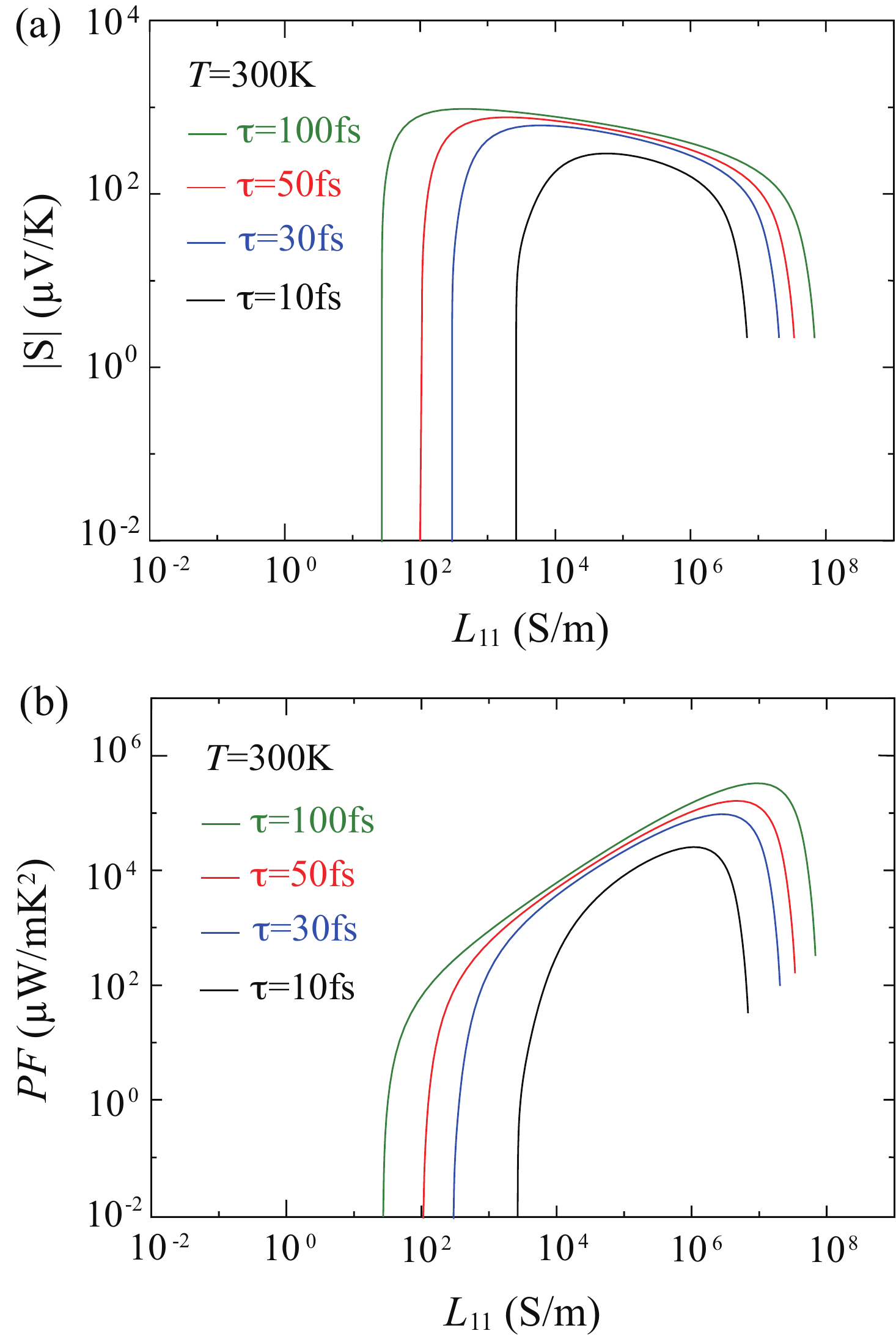}
  \end{center}
\caption{(a) $S$-$L_{11}$ plot and (b) $PF$-$L_{11}$ plot for (10,0) SWCNTs at $T=300$~K for $\tau=10$ (black curve), $30$ (blue curve), $50$ 
(red curve), and $100$~fs (green curve) calculated using the constant-$\tau$ approximation.}
\label{fig:S-L11_tau}
\end{figure}

Here, we show $S$ and $PF$ as a function of $L_{11}$ ({\it i.e.}, the $S$-$L_{11}$ plot and the $PF$-$L_{11}$ plot) for (10,0) SWCNTs at $T=300$~K for 
$\tau=10$ (black curve), $30$ (blue curve), $50$ (red curve), and $100$~fs (green curve) in Fig.~\ref{fig:S-L11_tau}(a) and \ref{fig:S-L11_tau}(b).

\subsection{Self-consistent Born approximation}
In this subsection, another refined approximation, {\it i.e.}, the self-consistent Born approximation (SCBA), is adopted for self-energy corrections
due to a disorder potential. For the disorder potential, a short-range random impurity potential is considered. It 
includes the following two types of scatterers,
\begin{eqnarray}
U(x)=U_1a\sum_{\langle{j}\rangle}^{N_1}\delta(x-x_j)
+U_2a\sum_{\langle{l}\rangle}^{N_2}\delta(x-x_l),
\label{eq:short-range_Ux}
\end{eqnarray}
where $x_j$ and $x_l$ represent the locations of different types of impurity, such as n-type and p-type impurities,
and $\langle{j}\rangle$ and $\langle{l}\rangle$ denote the sums of different impurity sites ({\it i.e.}, $j\neq l$).
The potential strengths $U_1$ and $U_2$ are given by
\begin{eqnarray}
U_1\equiv \left(
\begin{matrix}
u_{11} & 0 \\
0 & 0 \\
\end{matrix}
\right)\quad{\rm and}\quad
U_2\equiv
\left(
\begin{matrix}
0 & 0 \\
0 & u_{22} \\
\end{matrix}
\right).
\label{eq:U_example}
\end{eqnarray}
When the potential is expressed as Eq.~(\ref{eq:U_example}), the vertex correction vanishes, as shown in Appendix~\ref{sec:C}. 
In this case, the impurity potential in the momentum space is given by
\begin{eqnarray}
U(q)&=&\frac{1}{L}\int_{-\infty}^\infty \!\!\! dx~e^{-iqx} U(x)\\
&=&\frac{U_1}{N}\sum_{\langle{j}\rangle}e^{-iqx_j}+\frac{U_2}{N}\sum_{\langle{l}\rangle}e^{-iqx_l},
\label{eq:short-range_Uq}
\end{eqnarray}
where $N$ is the total number of unit cells in the system. 

\begin{figure}[t]
  \begin{center}
  \includegraphics[keepaspectratio=true,width=70mm]{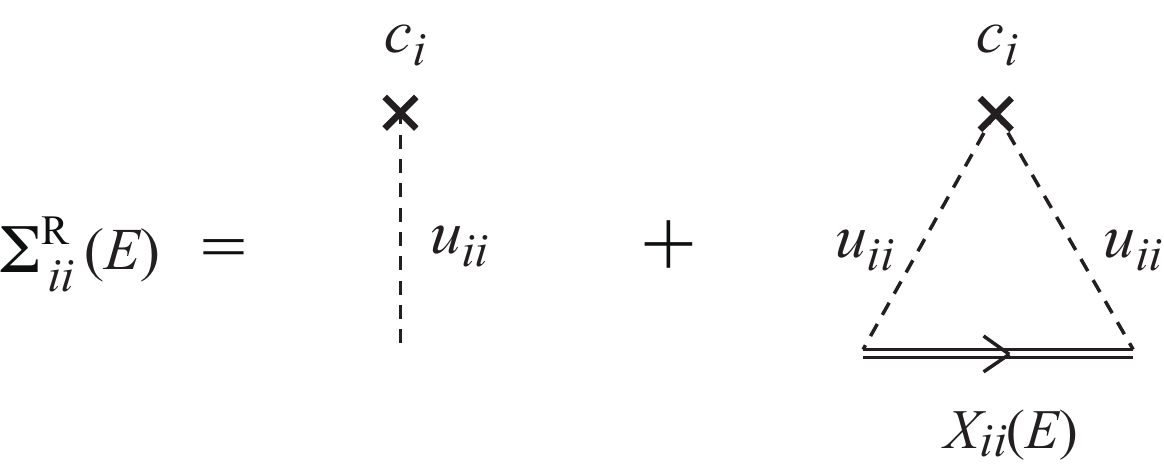}
  \end{center}
\caption{Self-consistent Born approximation for the retarded self-energy $\Sigma^{\rm R}_{ii}(E)$ ($i=1,2$) 
of a one-particle retarded Green's function. The $\times$ marks, dotted lines, and solid double lines with 
an arrow denote the impurity sites, impurity potential, and $k$-averaged retarded Green's function $X_{ii}(E)$ 
to be determined self-consistently, respectively.}
\label{fig:06}
\end{figure}

In the SCBA for the short-range potential in Eq.~(\ref{eq:short-range_Uq}) 
with Eq.~(\ref{eq:U_example}), the retarded self-energy matrix is given in a diagonal and $k$-independent form as
\begin{eqnarray}
\Sigma^{\rm R}(E)=\left(
\begin{matrix}
\Sigma^{\rm R}_{11}(E) & 0 \\
0 & \Sigma^{\rm R}_{22}(E) \\
\end{matrix}
\right), \quad {\rm Im}\Sigma^{\rm R}_{jj}(E)<0
\label{eq:diagonal_Sigma}
\end{eqnarray}
and the self-consistency equations are
\begin{eqnarray}
\Sigma^{\rm R}_{11}(E)=c_1u_{11}+c_1u_{11}^2X_{11}(E)
\label{eq:sigma_11}\\
\Sigma^{\rm R}_{22}(E)=c_2u_{22}+c_2u_{22}^2X_{22}(E)
\label{eq:sigma_22}
\end{eqnarray}
with $X_{jj}(E)\equiv \frac{1}{N}\sum_kG_{jj}^{\rm R}(k,E)$ ($j=1$, $2$). In Eqs.~(\ref{eq:sigma_11}) and
(\ref{eq:sigma_22}), $c_1\equiv N_1/N$ and $c_2\equiv N_2/N$ are the concentrations of impurities ({\it i.e.}, the impurity 
density per unit cell) with potential strengths $u_{11}$ and $u_{22}$, respectively (see Fig.~\ref{fig:06}). 
Moreover, $X_{11}(E)$ and $X_{22}(E)$ can be analytically calculated for the 1D Dirac electrons as
\begin{eqnarray}
X_{11}(E)&=&-i\frac{a}{2\hbar v}\frac{\kappa_2}{\sqrt{\kappa_1\kappa_2}},\\
X_{22}(E)&=&-i\frac{a}{2\hbar v}\frac{\kappa_1}{\sqrt{\kappa_1\kappa_2}}
\end{eqnarray}
with $\kappa_1\equiv (E-\Delta-\Sigma_{11}^{\rm R}(E))/\hbar v$, $\kappa_{2}\equiv (E+\Delta-\Sigma_{22}^{\rm R}(E))/\hbar v$,
and ${\rm Im}\sqrt{\kappa_1\kappa_2}>0$. Thus, the self-consistent equations for 1D Dirac electrons are given by
\begin{eqnarray}
\Sigma^{\rm R}_{11}(E)-c_1u_{11}=-i\frac{\hbar}{2\tau_1}\frac{\kappa_2}{\sqrt{\kappa_1\kappa_2}},
\label{eq:sigma_R_11}\\
\Sigma^{\rm R}_{22}(E)-c_2u_{22}=-i\frac{\hbar}{2\tau_2}\frac{\kappa_1}{\sqrt{\kappa_1\kappa_2}},
\label{eq:sigma_R_22}
\end{eqnarray}
where $\tau_1$ and $\tau_2$ are the relaxation times related to $u_{11}$ and $u_{22}$ in the limit $|E|\to\infty$,
which are defined as $\tau_1\equiv \frac{\hbar^2v}{c_1u_{11}^2a}$ and $\tau_2\equiv \frac{\hbar^2v}{c_2u_{22}^2a}$,
respectively.

\begin{figure}[t]
  \begin{center}
  \includegraphics[keepaspectratio=true,width=80mm]{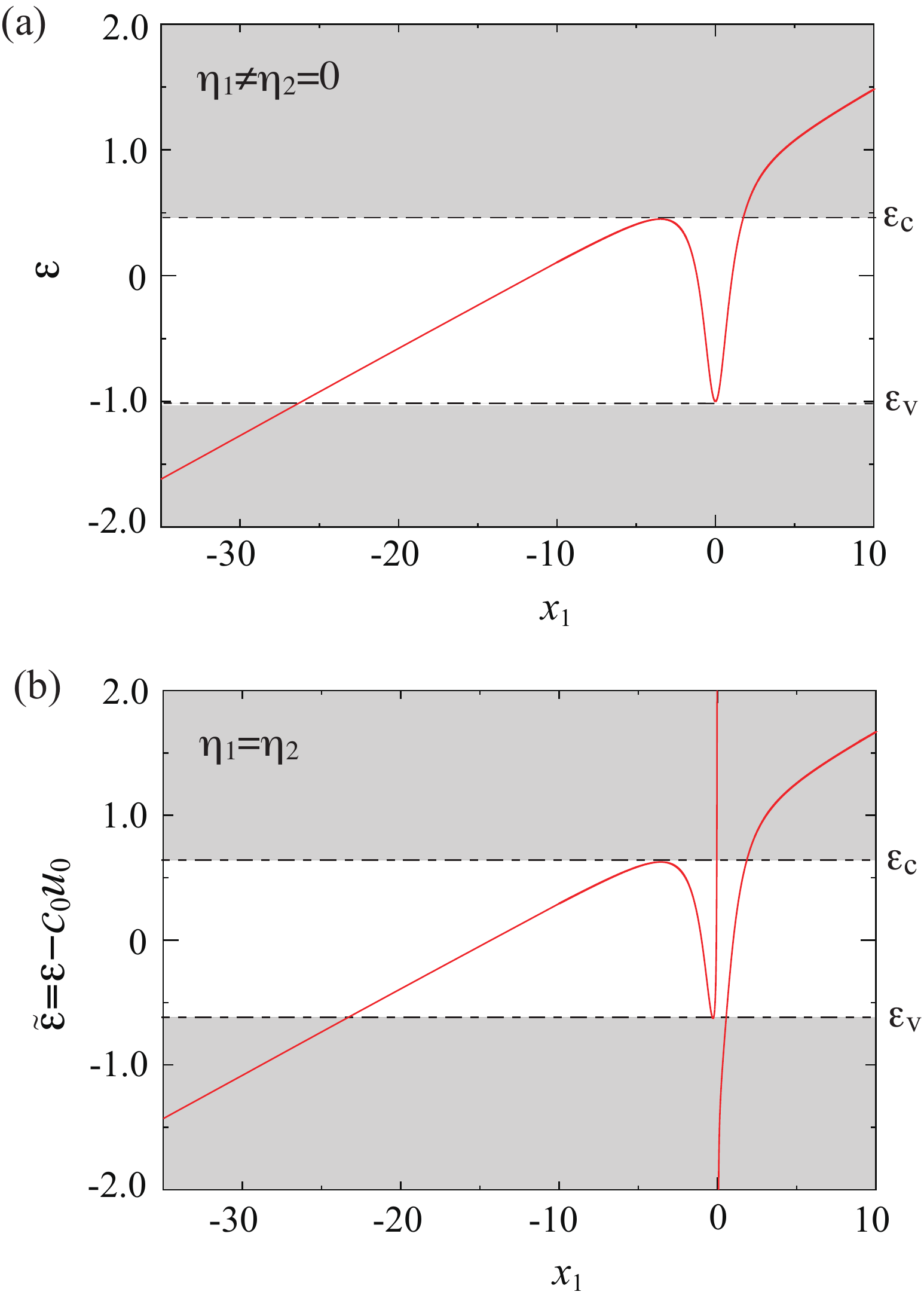}
  \end{center}
\caption{(Color online) 
$\epsilon$-$x_1$ relations for (a) $c_1=0.09$, $c_2=0$, and $u_{11}=-1.0$~eV ({\it i.e.}, $\eta_1=0.07$ and $\eta_2=0$ as well as
$\tau_1=10$~fs and $\tau_2=\infty$) and (b) $c_0(=c_1=c_2)=0.09$ and $u_0(=u_{11}=u_{22})=-1.0$~eV ({\it i.e.}, $\eta_1=\eta_2=0.07$ 
and $\tau_1=\tau_2=10$~fs). The shaded regions indicate the energy regions where the DOS is finite. The broken lines denote the mobility 
edges $\epsilon_{\rm c}$ and $\epsilon_{\rm v}$.}
\label{fig:07}
\end{figure}

The simultaneous equations in Eqs.~(\ref{eq:sigma_R_11}) and (\ref{eq:sigma_R_22}) can be rewritten as 
the following equations with respect to $\sigma_1\equiv(\Sigma^{\rm R}_{11}(E)-c_1u_{11})/\Delta$ and
$\sigma_2\equiv(\Sigma^{\rm R}_{22}(E)-c_1u_{22})/\Delta$.
\begin{eqnarray}
\sigma_1^4-(\epsilon-1-c_1\tilde{u}_{11})\sigma_1^3-\eta_1^2(\epsilon+1-c_2\tilde{u}_{22})\sigma_1-\eta_1^3\eta_2=0
\label{eq:quartic_eq_1}
\end{eqnarray}
and $\sigma_1\sigma_2=-\eta_1\eta_2$ with $\epsilon\equiv E/\Delta$, $\eta_1\equiv \hbar/(2\Delta\tau_1)$, $\eta_2\equiv \hbar/(2\Delta\tau_1)$, 
$\tilde{u}_{11}\equiv{u_{11}/\Delta}$, and $\tilde{u}_{22}\equiv{u_{22}/\Delta}$. 
Equation (\ref{eq:quartic_eq_1}) indicates that for each $\epsilon$, there are four solutions of $\sigma_1(\epsilon)$, respectively: four real ones or two real and two complex ones,
the latter leading to finite DOS.
In order to clearly show the difference between the two cases, Eq.~(\ref{eq:quartic_eq_1}) is rewritten as 
\begin{eqnarray}
\epsilon=\frac{\eta_1x_1^3+(1+c_1\tilde{u}_{11})x_1^2-(1-c_2\tilde{u}_{22})}{x_1^2+1}-\frac{\eta_2}{x_1(x_1^2+1)},
\label{eq:e_eq_1_gen}
\end{eqnarray}
with $x_1\equiv \sigma_1/\eta_1$. 

In the special case of $\eta_2=0$, Eq.~(\ref{eq:quartic_eq_1}) gives a cubic equation with respect to real $x_1 (=\sigma_1/\eta_1)$. 
Figure~\ref{fig:07}(a) shows Eq.~(\ref{eq:e_eq_1_gen}) as a function of real $x_1$ for 
$\eta_1=0.07$ and $\eta_2=0$ for (10,0) SWCNTs with $2\Delta=0.95$~eV. In the shaded regions in Fig.~\ref{fig:07}(a), Eq.~(\ref{eq:quartic_eq_1}) 
has two complex and one real solutions of $x_1$. DOS is finite in these energy regions. The boundaries between the finite- and zero-DOS regions 
($\epsilon_{\rm c}$ and $\epsilon_{\rm v}$ in Fig.~\ref{fig:07}(a)), which are band edges, can be determined using the condition $d\epsilon/dx_1=0$. 
In coherent potential approximation (CPA) methods, including the present SCBA, the spectral conductivity $\alpha(E)$ becomes finite once DOS becomes finite (see \S~\ref{sec:3.2.1}), 
since CPA ignores the effects of Anderson localization due to the interference effects of scattered waves, which can lead to finite DOS
even in the energy region where the conductivity is zero. It is known that every state is localized in one and two dimensions in the presence of finite scattering~\cite{rf:nagaoka}.
However, once the system size or temperature becomes finite, the effects of Anderson localization are greatly reduced. This situation is assumed in the present study 
and hence the band edges in the CPA are used to represent the effective mobility edges. For the case of $\eta_2=0$, one of the mobility edges is always
$\epsilon_{\rm v}=-1$ ({\it i.e.}, $E_{\rm v}=-\Delta$), as shown in Fig.~\ref{fig:07}(a).

The case of $c_1=c_2(\equiv{c_0})$ which we call symmetric case, $\tilde{u}_{11}=\tilde{u}_{22}(\equiv \tilde{u}_0)$ ({\it i.e.}, 
$\eta_1=\eta_2(\equiv\eta_0)$) is also considered. For this case, Eq.~(\ref{eq:e_eq_1_gen}) becomes
\begin{eqnarray}
\tilde{\epsilon}=\frac{\eta_0x_1^3+x_1^2-1}{x_1^2+1}-\frac{\eta_0}{x_1(x_1^2+1)},
\label{eq:e_eq_1}
\end{eqnarray}
with $\tilde{\epsilon}\equiv \epsilon-c_0\tilde{u}_0$. 
Figure~\ref{fig:07}(b) shows $\tilde{\epsilon}$ in Eq.~(\ref{eq:e_eq_1}) as a function of real $x_1$ for $\eta_1=\eta_2=0.07$, corresponding to 
(10,0) SWCNTs with $2\Delta=0.95$~eV, and $\tau_1=\tau_2=10$~fs. In the shaded regions in Fig.~\ref{fig:07}(b), Eq.~(\ref{eq:quartic_eq_1}) 
has two complex and two real solutions of $x_1$. These regions have finite DOS. The two mobility edges ($\epsilon_{\rm c}$ and $\epsilon_{\rm v}$) 
satisfy $\epsilon_{\rm c}=-\epsilon_{\rm v}$, as shown in Fig.~\ref{fig:07}(b).

\begin{figure}[t]
  \begin{center}
  \includegraphics[keepaspectratio=true,width=80mm]{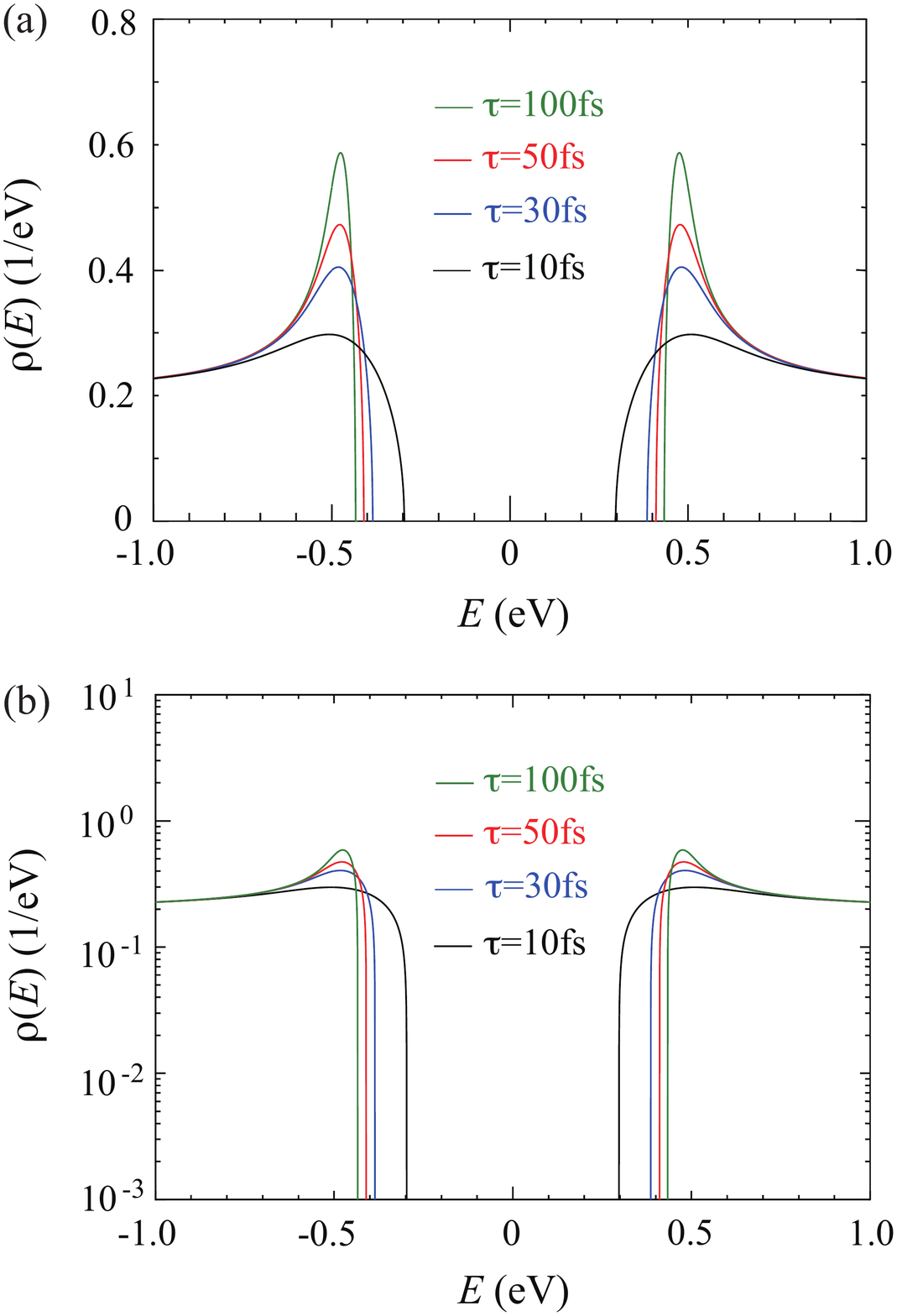}
  \end{center}
\caption{(Color online) (a) Density of states of the lowest-conduction and the highest-valence bands of (10,0) SWCNTs 
per spin and orbital for $\tau=10$ (black curve), $30$ (blue curve), $50$ (red curve), and $100$~fs (green curve) calculated using the SCBA and 
(b) the corresponding semi-log plots. }
\label{fig:08}
\end{figure}

The following discussion mainly  focuses on thermoelectric properties of SWCNTs for the symmetric case of $\eta_1=\eta_2$ ({\it i.e.}, $\tau_1=\tau_2$).
The thermoelectric properties for the asymmetric case of $\eta_1\neq\eta_2$ will be reported elsewhere.

\subsubsection{Density of states and spectral conductivity~\label{sec:3.2.1}}
Once the self energy $\Sigma^{\rm R}(E)$ is obtained via the above procedure, 
the DOS can be calculated using
\begin{eqnarray}
\rho(E)&=&-\frac{1}{\pi}\sum_{j=1,2}{\rm Im}X_{jj}(E)\nonumber\\
&=&\frac{a}{2\pi\hbar v}{\rm Re}\left\{\frac{\kappa_1+\kappa_2}{\sqrt{\kappa_1\kappa_2}}\right\}.
\label{eq:dos_dif}
\label{eq:DOS}
\end{eqnarray}
Figure~\ref{fig:08}(a) shows the calculated DOS values near the band gap of (10,0) SWCNTs 
per spin and orbital for $\tau(\equiv\tau_1=\tau_2)=10$ (black curve), $30$ (blue curve), $50$ (red curve), and $100$~fs (green curve) 
and Fig.~\ref{fig:08}(b) shows the corresponding semi-log plots. In contrast to the DOS values calculated using the constant-$\tau$ approximation (see Fig.~\ref{fig:01}), 
clear mobility edges ({\it i.e.}, $E_{\rm c}$ and $E_{\rm v}=-|E_{\rm v}|$) exist, as shown in Fig.~\ref{fig:08}. As $\tau$ decreases, 
the band gap becomes small and the value of the DOS peak decreases. It can also be seen that the DOS near the mobility edges shows the behaviors 
$\rho(E)\propto\sqrt{E-E_{\rm c}}$ for $E\ge E_{\rm c}(>0)$ and $\rho(E)\propto\sqrt{E_{\rm v}-E}$ for $E\le E_{\rm v}(<0)$, respectively. 

\begin{figure}[t]
  \begin{center}
  \includegraphics[keepaspectratio=true,width=80mm]{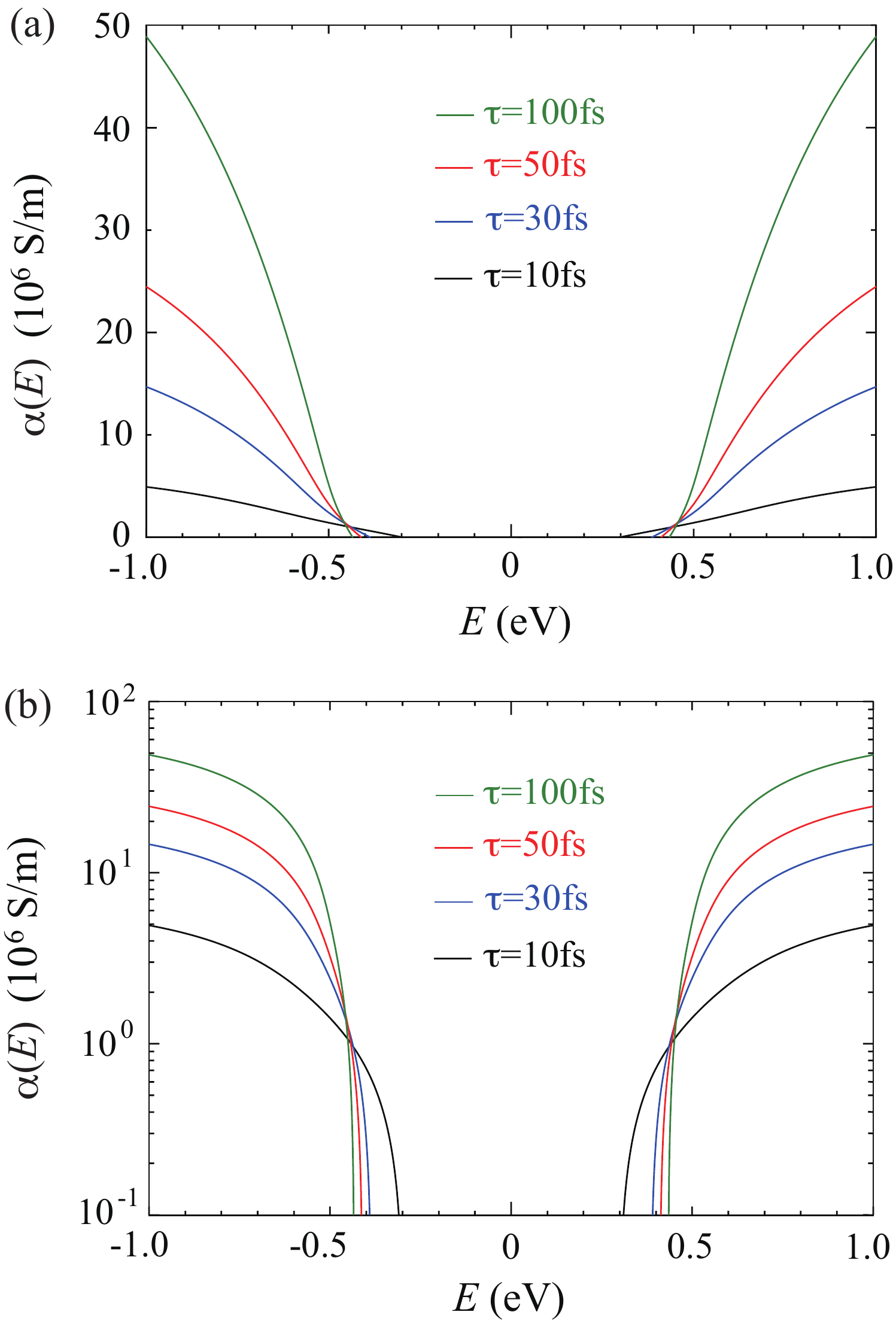}
  \end{center}
\caption{(Color online) (a) Spectral conductivities of (10,0) SWCNTs for $\tau=10$ (black curve), $30$ (blue curve), $50$ (red curve), 
and $100$~fs (green curve) calculated using the SCBA and (b) the corresponding semi-log plots. }
\label{fig:09}
\end{figure}

When the self-energy matrix is given in a diagonal and $k$-independent form, as shown in Eq.~(\ref{eq:diagonal_Sigma}), 
the spectral conductivity $\alpha(E)$ in Eq.~(\ref{eq:spectral_conductivity}) can be analytically calculated as
\begin{eqnarray}
\alpha(E)=\frac{1}{A}\frac{e^2}{h}\frac{1}{{\rm Im}(\kappa_1\kappa_2)}
{\rm Re}\left\{
\frac{2\kappa_1\kappa_2+\kappa_1^*\kappa_2+\kappa_1\kappa_2^*}{\sqrt{\kappa_1\kappa_2}}
\right\}
\label{eq:spectral_conductivity_ana}
\end{eqnarray}
with ${\rm Im}\sqrt{\kappa_1\kappa_2}>0$. Note that Eq.~(\ref{eq:spectral_conductivity_ana}) reduces to Eq.~(\ref{eq:const-tau_alpha}) 
in the constant-$\tau$ approximation ({\it i.e.}, $\Sigma^{{\rm R}}=-{i}\hbar/2\tau$).
Figure~\ref{fig:09}(a) shows the spectral conductivity $\alpha(E)$ of (10,0) SWCNTs for $\tau(\equiv\tau_1=\tau_2)=10$ (black curve), 
$30$ (blue curve), $50$ (red curve), and $100$~fs (green curve) and 
Fig.~\ref{fig:09}(b) shows the corresponding semi-log plots. Here, $\alpha(E)$ in Eq.~(\ref{eq:spectral_conductivity_ana}) was multiplied by a factor of 4. 
In contrast to the results obtained with the constant-$\tau$ approximation (see Fig.~\ref{fig:02}), 
$\alpha(E)$ has a clear gap, as shown in Figs.~\ref{fig:09}(a) and \ref{fig:09}(b). It can also be seen that $\alpha(E)$ near the mobility edges $E=E_{\rm c}(>0)$ and $E=E_{\rm v}(<0)$
shows the behaviors $\alpha(E)\propto (E-E_{\rm c})$ for $E\ge E_{\rm c}$ and $\alpha(E)\propto (E_{\rm v}-E)$ for $E\le E_{\rm v}$, respectively 
(see Appendix~\ref{sec:D} for details).

\begin{figure}[t]
  \begin{center}
  \includegraphics[keepaspectratio=true,width=80mm]{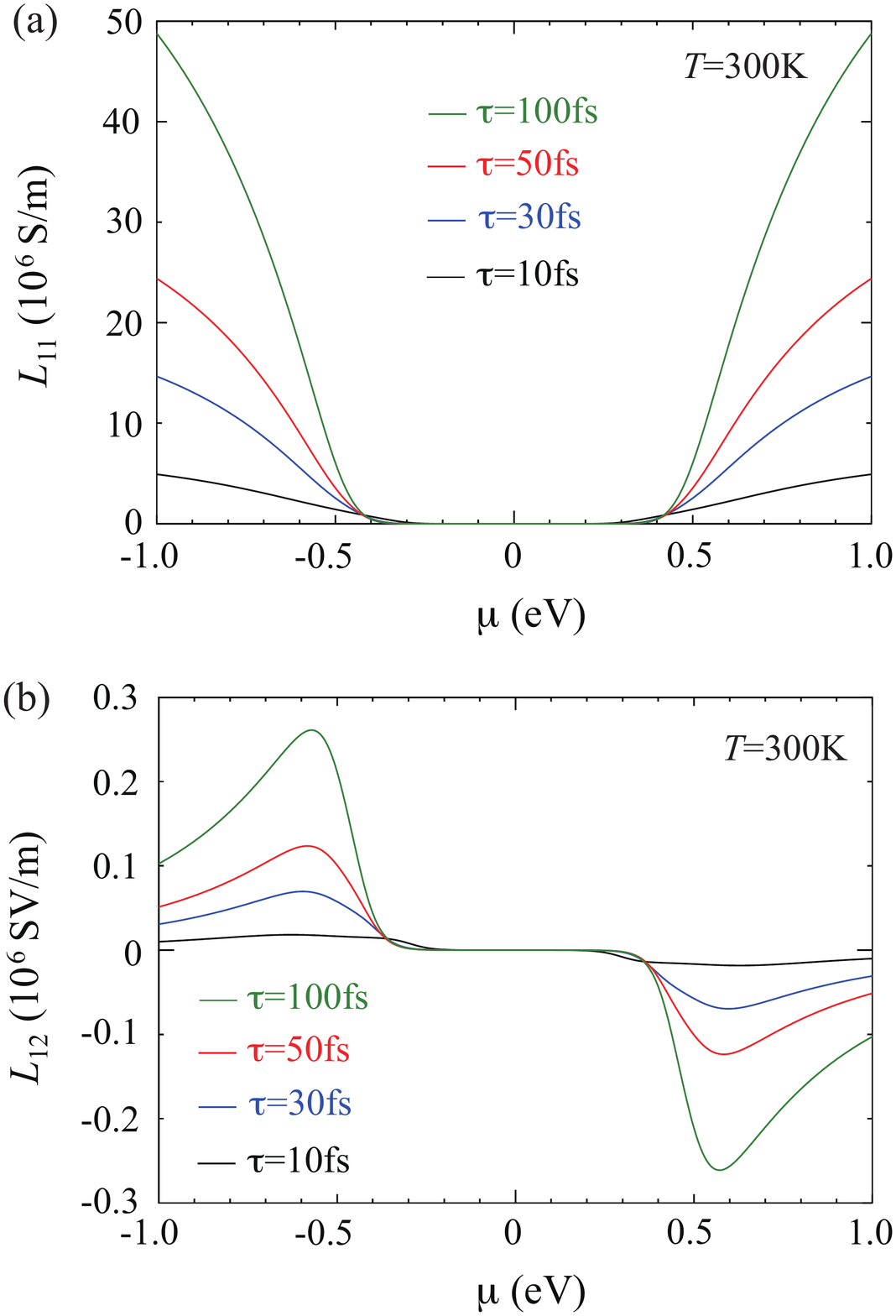}
  \end{center}
\caption{(Color online) Chemical potential dependence of (a) electrical conductivity $L_{11}$ and (b) thermoelectrical conductivity $L_{12}$ 
of (10,0) SWCNTs at 300~K for $\tau=10$ (black curve), $30$ (blue curve), $50$ (red curve), and $100$~fs (green curve) calculated using the SCBA. }
\label{fig:10}
\end{figure}

\subsubsection{Chemical potential dependence of $L_{11}$ and $L_{12}$ at 300~K}
Figure~\ref{fig:10}(a) shows the $\mu$ dependence of the $L_{11}$ of (10,0) SWCNTs at 300~K for 
$\tau(\equiv\tau_1=\tau_2)=10$ (black curve), $30$ (blue curve), $50$ (red curve), and $100$~fs (green curve). 
As expected from Eq.~(\ref{eq:L_11}), the $\mu$ dependence of $L_{11}$ for each $\tau$ shows features similar to those for 
the $E$ dependence of the spectral conductivity $\alpha(E)$ in Fig.~(\ref{fig:08}). 
As shown in Fig.~\ref{fig:10}(b), the $L_{12}$ values of (10,0) SWCNTs for $\tau=10$, $30$, $50$, and $100$~fs at $T=300$~K 
have peaks and dips near the mobility edges ($E=E_{\rm c}$ and $E_{\rm v} (=-E_{\rm c})$), respectively. 
These characteristics of $L_{11}$ and $L_{12}$ are essentially the same as those obtained using the constant-$\tau$ approximation in the previous section.

\begin{figure}[t]
  \begin{center}
  \includegraphics[keepaspectratio=true,width=80mm]{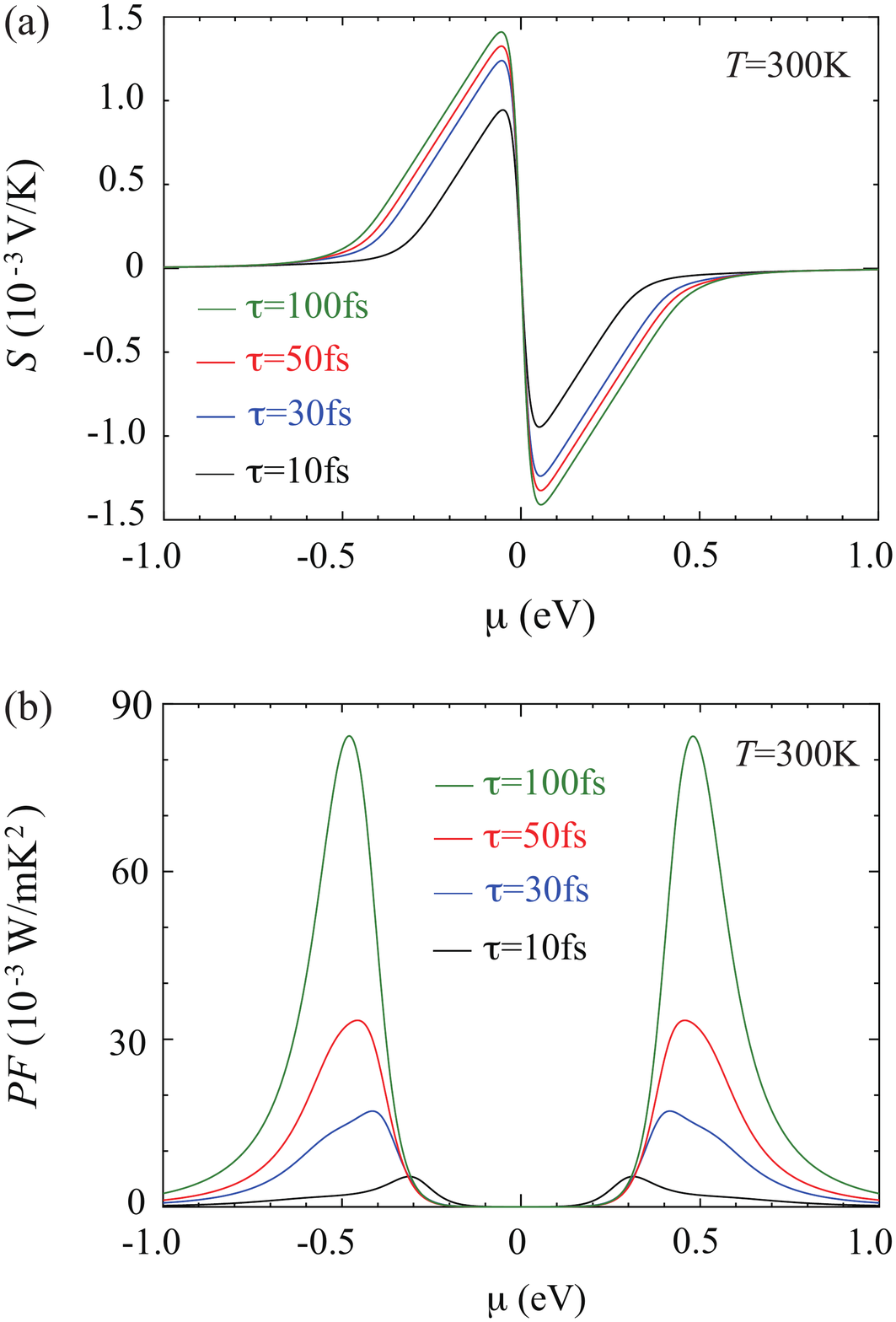}
  \end{center}
\caption{(Color online) Chemical potential dependence of (a) Seebeck coefficient $S$ and (b) power factor $PF$ of (10,0) SWCNTs 
at 300 K for $\tau=10$ (black curve), $30$ (blue curve), $50$ (red curve), and $100$~fs (green curve) calculated using the SCBA. }
\label{fig:11}
\end{figure}

\subsubsection{Chemical potential dependence of $S$ and $PF$ at 300 K}
Figure~\ref{fig:11}(a) shows the $\mu$ dependence of the Seebeck coefficient $S$ of (10,0) SWCNTs for $\tau=10$, $30$, $50$, and $100$~fs at 300 K, 
which was obtained by substituting the $L_{11}$ and $L_{12}$ data in Fig.~\ref{fig:10} into Eq.~(\ref{eq:S}). As shown in Fig.~\ref{fig:11}(a), the absolute 
value of $S$ decreases with decreasing $\tau$, and $S$ exhibits bipolar effects as a function of $\mu$ ({\it i.e.}, a bipolar thermoelectric effect)~\cite{rf:yanagi,rf:Shimizu}. 
In contrast to the results shown in Fig.~\ref{fig:04}(a), the optimal chemical potentials $\mu=\pm\mu_{\rm opt}$ are almost independent of $\tau$.
Figure~\ref{fig:11}(b) shows the power factor $PF$ of (10,0) SWCNTs for $\tau=10$, $30$, $50$, and $100$~fs at $T=300$~K,
which was calculated by substituting the $L_{11}$ and $L_{12}$ data in Fig.~\ref{fig:10} into Eq.~(\ref{eq:PF}). Similar to the results obtained using
the constant-$\tau$ approximation, $PF$ has a maximum value near the conduction and valence band edges of pristine (10,0) SWCNTs.

\begin{figure}[t]
  \begin{center}
  \includegraphics[keepaspectratio=true,width=80mm]{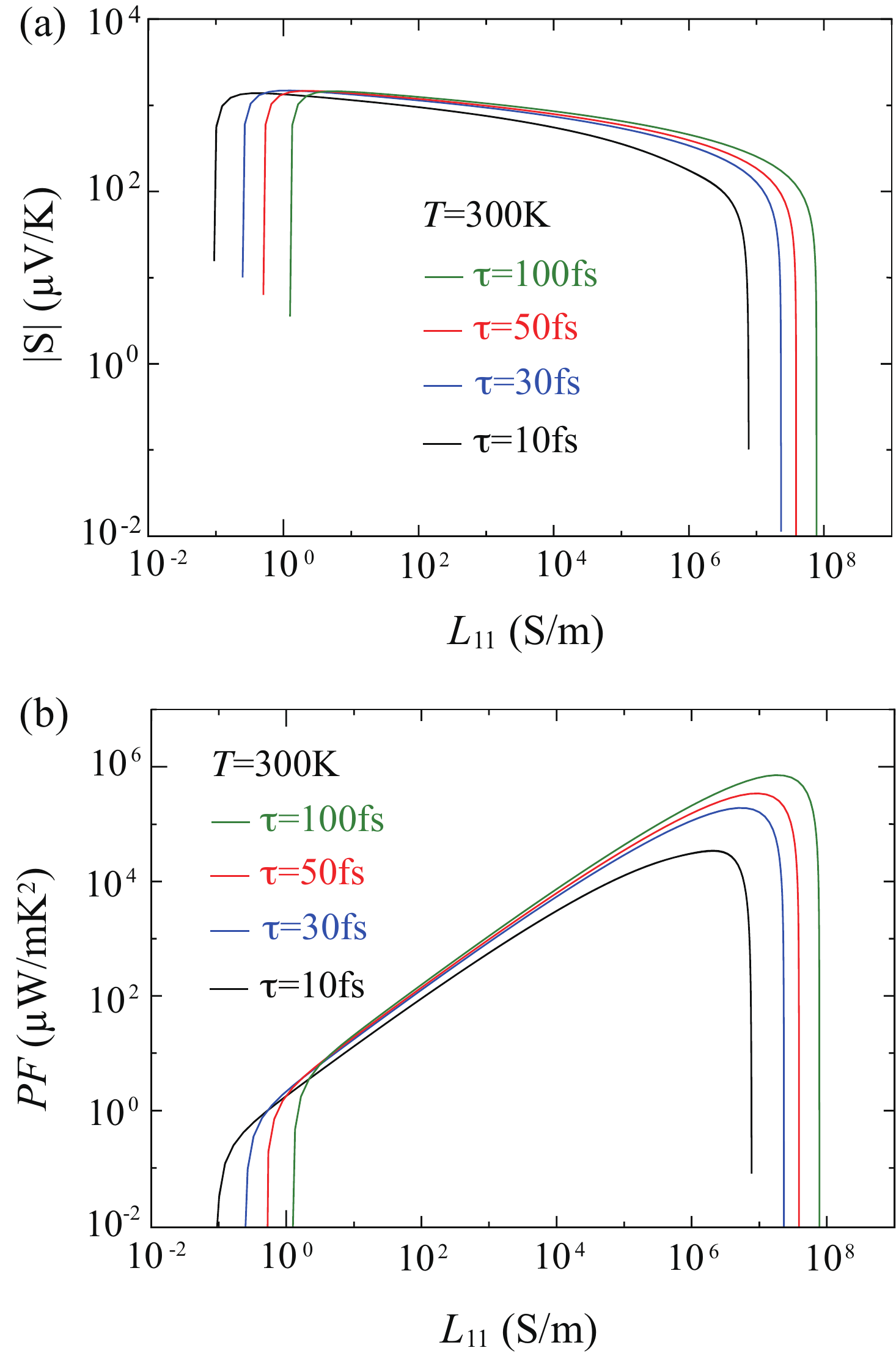}
  \end{center}
\caption{(a) $S$-$L_{11}$ plot and (b) $PF$-$L_{11}$ plot for (10,0) SWCNTs at $T=300$~K for $\tau=10$ (black curve), $30$ (blue curve), $50$ 
(red curve), and $100$~fs (green curve) calculated using the SCBA.}
\label{fig:S-L11_SCBA}
\end{figure}

Here, we show $S$ and $PF$ as a function of $L_{11}$ ({\it i.e.}, the $S$-$L_{11}$ plot and the $PF$-$L_{11}$ plot) for (10,0) SWCNTs at $T=300$~K for 
$\tau=10$ (black curve), $30$ (blue curve), $50$ (red curve), and $100$~fs (green curve) in Fig.~\ref{fig:S-L11_SCBA}(a) and \ref{fig:S-L11_SCBA}(b).

\subsection{Low-temperature behavior of Seebeck coefficient}
In this section, we discuss the low-temperature behavior of $S$ for SWCNTs within SCBA.
Since a clear band gap exists, as shown in Fig.~\ref{fig:09}, the spectral conductivity $\alpha(E)$ can be divided into two parts as
\begin{eqnarray}
\alpha(E)=\alpha_{\rm e}(E)\theta(E-E_{\rm c})+\alpha_{\rm h}(E)\theta(E-E_{\rm v}),
\label{eq:alpha_eh}
\end{eqnarray}
where $\alpha_{\rm e(h)}(E)$ is the conduction-electron (valence-hole) spectral conductivity and $\theta(x)$ is the Heaviside step function.
 For this case, the Seebeck coefficient $S$ can be rewritten as
\begin{eqnarray}
S(E)=\frac{1}{T}\frac{L_{12}^{\rm e}+L_{12}^{\rm h}}{L_{11}^{\rm e}+L_{11}^{\rm h}}=
\frac{L_{11}^{\rm e}S^{\rm e}+L_{11}^{\rm h}S^{\rm h}}{L_{11}^{\rm e}+L_{11}^{\rm h}}
\label{eq:S_eh}
\end{eqnarray}
with
\begin{eqnarray}
L_{11}^{\rm e}&\equiv& \int_{E_{\rm c}}^\infty \!\!\! dE \left(-\frac{\partial f(E-\mu)}{\partial E}\right)\alpha_{\rm e}(E)\\
\label{eq:L11_e}
L_{11}^{\rm h}&\equiv& \int_{-\infty}^{E_{\rm v}} \!\!\! dE \left(-\frac{\partial f(E-\mu)}{\partial E}\right)\alpha_{\rm h}(E)\\
\label{eq:L11_h}
L_{12}^{\rm e}&\equiv& -\frac{1}{e}\int_{E_{\rm c}}^\infty \!\!\! dE \left(-\frac{\partial f(E-\mu)}{\partial E}\right)(E-\mu)\alpha_{\rm e}(E)\\
\label{eq:L12_e}
L_{12}^{\rm h}&\equiv& -\frac{1}{e}\int_{-\infty}^{E_{\rm v}} \!\!\! dE \left(-\frac{\partial f(E-\mu)}{\partial E}\right)(E-\mu)\alpha_{\rm h}(E)
\label{eq:L12_h}
\end{eqnarray}
and $S^{\rm e(h)}\equiv L_{12}^{\rm e(h)}/TL_{11}^{\rm e(h)}$. Here, the superscripts e and h represent electrons and holes, respectively.
Note that although Eq.~(\ref{eq:S_eh}) is formally the same as
the TIB model for the Seebeck coefficient based on the Boltzmann transport theory, Eqs.~(\ref{eq:L11_e})-(\ref{eq:L12_h}) include the effects of 
inter-band scattering between the conduction and valence bands, which is not taken into account in TIB model.

\begin{figure}[t]
  \begin{center}
  \includegraphics[keepaspectratio=true,width=80mm]{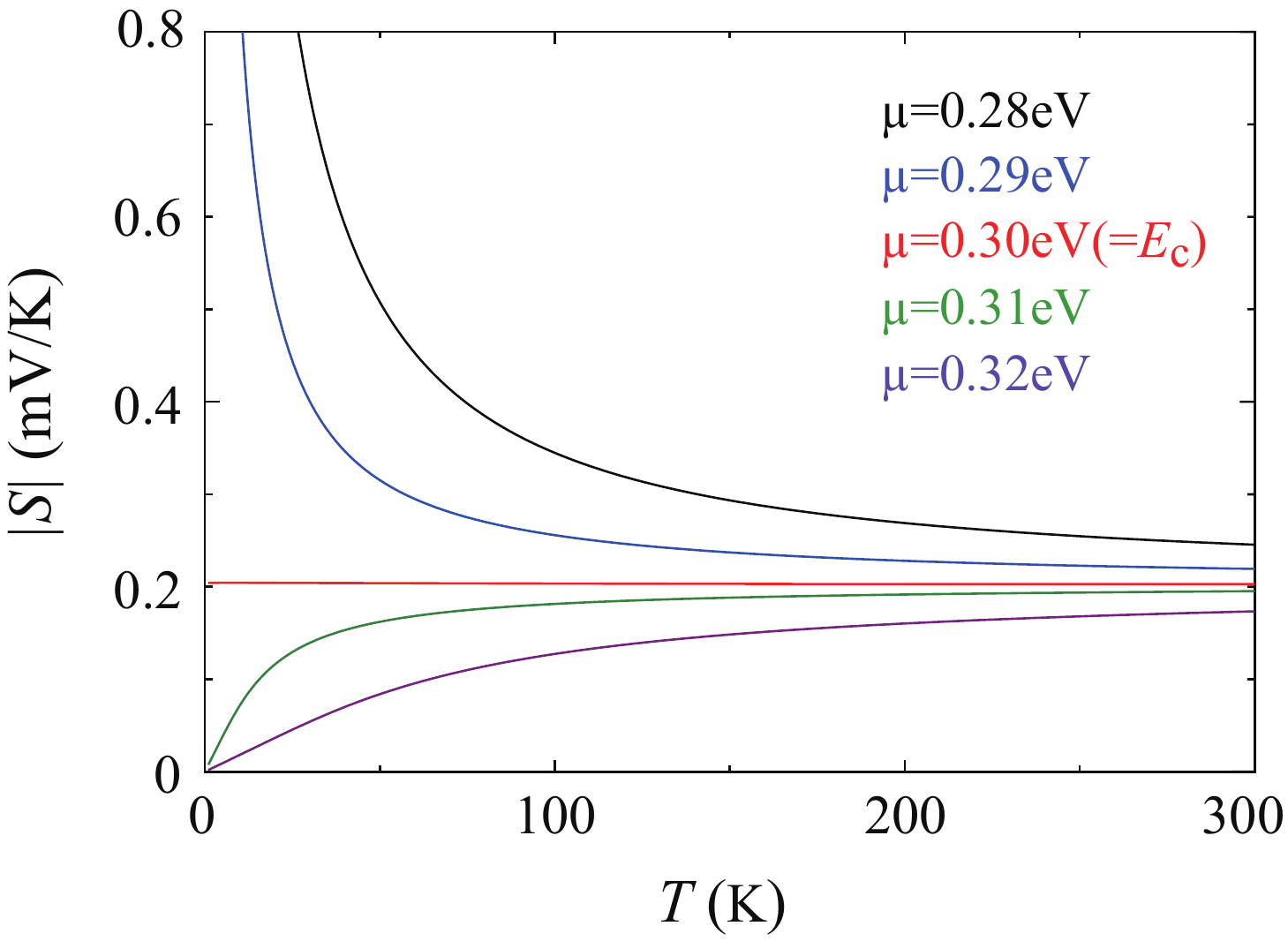}
  \end{center}
\caption{(Color online) Temperature dependence of the Seebeck coefficient of (10,0) SWCNTs with $\tau=10$~fs for 
$\mu=0.28$~eV (black curve), $0.29$~eV (blue curve), $0.30$eV ($=E_{\rm c}$) (red curve), $0.31$~eV (green curve), and $0.32$~eV (purple curve).
Here, $E_{\rm c}$ is the mobility edge.}
\label{fig:13}
\end{figure}

The temperature dependence of the Seebeck coefficient for a symmetric case satisfying $E_{\rm v}=-E_{\rm c}=\Delta$ is now discussed. 
For the case of $\mu>E_{\rm c}$ and $2\Delta\gg k_{\rm B}T$, $L_{11}^{\rm h}$ and $L_{12}^{\rm h}$ can be neglected and 
$S\approx S^{\rm e}$.
For this case, $S^{\rm e}$ in the low-temperature limit can be easily obtained as
\begin{eqnarray}
S^{\rm e}\approx-\frac{\pi^2k_{\rm B}^2T}{3e}\left(\frac{d\ln\alpha_{\rm e}(E)}{dE}\right)_{E=\mu},\quad (\mu>E_{\rm c})
\label{eq:mott}
\end{eqnarray}
by performing the Sommerfeld expansion of Eqs.~(\ref{eq:L11_e})-(\ref{eq:L12_h}).
The $T$-linear behavior of $S$, known as Mott's formula~\cite{rf:mott}, can be seen for the (10,0) SWCNTs 
when $\mu$ is larger than $E_{\rm c}=0.30$~eV, as shown in Fig.~\ref{fig:13}. The $T$-linear region becomes smaller as $\mu$
approaches the mobility edge $E_{\rm c}$.

 When $\mu=E_{\rm c}$, the above-mentioned $T$-linear region vanishes and $S^{\rm e}$ becomes constant as
\begin{eqnarray}
S^{\rm e}=\frac{k_{\rm B}}{e}\frac{\int_{0}^{\infty}dx~\left(-\frac{\partial f(x)}{\partial x}\right)x^{2}}{\int_{0}^{\infty}\!\!\! dx~\left(-\frac{\partial f(x)}{\partial x}\right)x}
\approx \frac{k_{\rm B}}{e}\times{2.37},\quad (\mu=E_{\rm c}),
\label{eq:S_00}
\end{eqnarray}
where $f(x)=1/(1+e^x)$ with $x\equiv E-E_{\rm c}$. For $\mu=E_{\rm c}(=0.30~{\rm eV})$, the Seebeck coefficient $S$ of (10,0) SWCNTs with 
$\tau=10$~fs is constant with respect to $T$, as shown in Fig.~\ref{fig:13}.

When $\mu<E_{\rm c}$ but as far as $E_{\rm c}-\mu\ll 2\Delta$ together with at low temperature $k_{\rm B}T\ll E_{\rm c}-\mu$, 
the valence holes are frozen out ({\it i.e.}, $L_{11}^{\rm h}=0$ and $L_{12}^{\rm h}=0$)
and the Seebeck coefficient $S\approx S^{\rm e}$, which is inversely proportional to $T$ as
\begin{eqnarray}
S^{\rm e}\approx -S_0-\frac{E_{\rm c}-\mu}{e T},\quad (\mu< E_{\rm c};~k_{\rm B}T\ll E_{\rm c}-\mu\ll 2\Delta).
\label{eq:S_T-1}
\end{eqnarray}
with
\begin{eqnarray}
S_0=\frac{k_{\rm B}}{e}\frac{\int_{0}^{\infty}dx~e^xx^{2}}{\int_{0}^{\infty}\!\!\! dx~e^xx}=\frac{2k_{\rm B}}{e}.
\label{eq:S_00}
\end{eqnarray}
Here, we note $f(E-\mu)\approx e^{-(E-\mu)/k_{\rm B}T}$ for $E>E_{\rm c}$ and
$\alpha(E)\propto(E-E_{\rm c})$ near $E=E_{\rm c}$.
The $T^{-1}$ behavior of $S$ can be seen for (10,0) SWCNTs with 
$\tau=10$~fs for $\mu=0.28$~eV and $0.29$~eV$(<E_{\rm c}=0.30$~eV) in Fig.~\ref{fig:13}. 

\section{Comparison with experiments}
As shown in Figs.~\ref{fig:04}(a) and \ref{fig:11}(a), the present theory naturally leads to the bipolar thermoelectric effects of SWCNTs, {\it i.e.}, 
the sign inversion of the Seebeck coefficient from positive (p-type) to negative (n-type) when the chemical potential is changed at room temperature
as observed in experiments~\cite{rf:yanagi,rf:Shimizu}.

Regarding the tradeoff relation between $S$ and $L_{11}$ for SWCNTs at room temperature
{\it i.e.}, decreasing $S$ with increasing $L_{11}$ as seen in experiment~\cite{rf:macLeod} would be understood by our theoretical results in 
Fig.~\ref{fig:S-L11_tau} and Fig.~\ref{fig:S-L11_SCBA}.
It is to be noted that the $S$-$L_{11}$ tradeoff relation changes when $\mu$ is located near the middle of the band gap, which
can be observed in experiments using the FET setup.
This tradeoff between $S$ and $L_{11}$ would be naturally reflected in between $PF$ and $L_{11}$ as indicated in Fig.~5(b) and Fig.~12(b).

In addition, the crossover from $T$-linear to $T$-inverse behavior of the Seebeck coefficient at low temperature 
in Fig.~\ref{fig:13} will also be an interesting experimental challenge.

\section{Conclusion}
The present study based on 1D Dirac electrons has developed a theoretical framework of bipolar thermoelectric effects in SWCNT described as 1d Dirac electrons under disorder. 
Based on the thermal Green's functions, effects of disorder have been treated within self-consistent Born approximation (SCBA), which is the simplest 
version of coherent potential approximation (CPA). The results has led to prediction of characteristic behaviors of Seebeck coefficient and power factor 
of semiconducting SWCNT, including the sign change of Seebeck coefficient as a function of chemical potential (gate voltage) as observed in recent experiments~\cite{rf:yanagi,rf:Shimizu}. It is to be noted that the effects of Anderson localization, which will play important roles at low temperatures, are totally ignored in the present study since our interest here in mainly at elevated temperatures, {\it e.g.}, room temperature.
We have also studied the crossover from $T$-linear to $T$-inverse behavior of the Seebeck coefficient
of semiconducting SWCNTs at low $T$ when the chemical potential changes from $\mu>E_{\rm c}$ to $\mu<E_{\rm c}$. 
The $S\propto{1/T}$ behavior for $\mu<E_{\rm c}$, which is commonly seen in text books~\cite{rf:MJ-text,rf:loffe}, should be taken with care 
for the temperature range of its observability because the present linear response theory breaks down in the limit of $T\to{0}$~\cite{rf:saso}, 
which needs separate and detailed studies.

\begin{acknowledgment}
The authors would like to thank Masao Ogata, Hiroyasu Matsuura, Hideaki Maehashi, Satoru Konabe, and 
Kenji Sasaoka for valuable discussions, and also Kazuhiro Yanagi for providing experimental data 
on the thermoelectric effects of carbon nanotubes. This work was supported, in part, by a 
JSPS KAKENHI grant (no. 15H03523).
\end{acknowledgment}

\appendix
\section{1D Dirac electrons in semiconducting carbon nanotubes~\label{sec:A}}

\begin{figure}[t]
  \begin{center}
  \includegraphics[keepaspectratio=true,width=80mm]{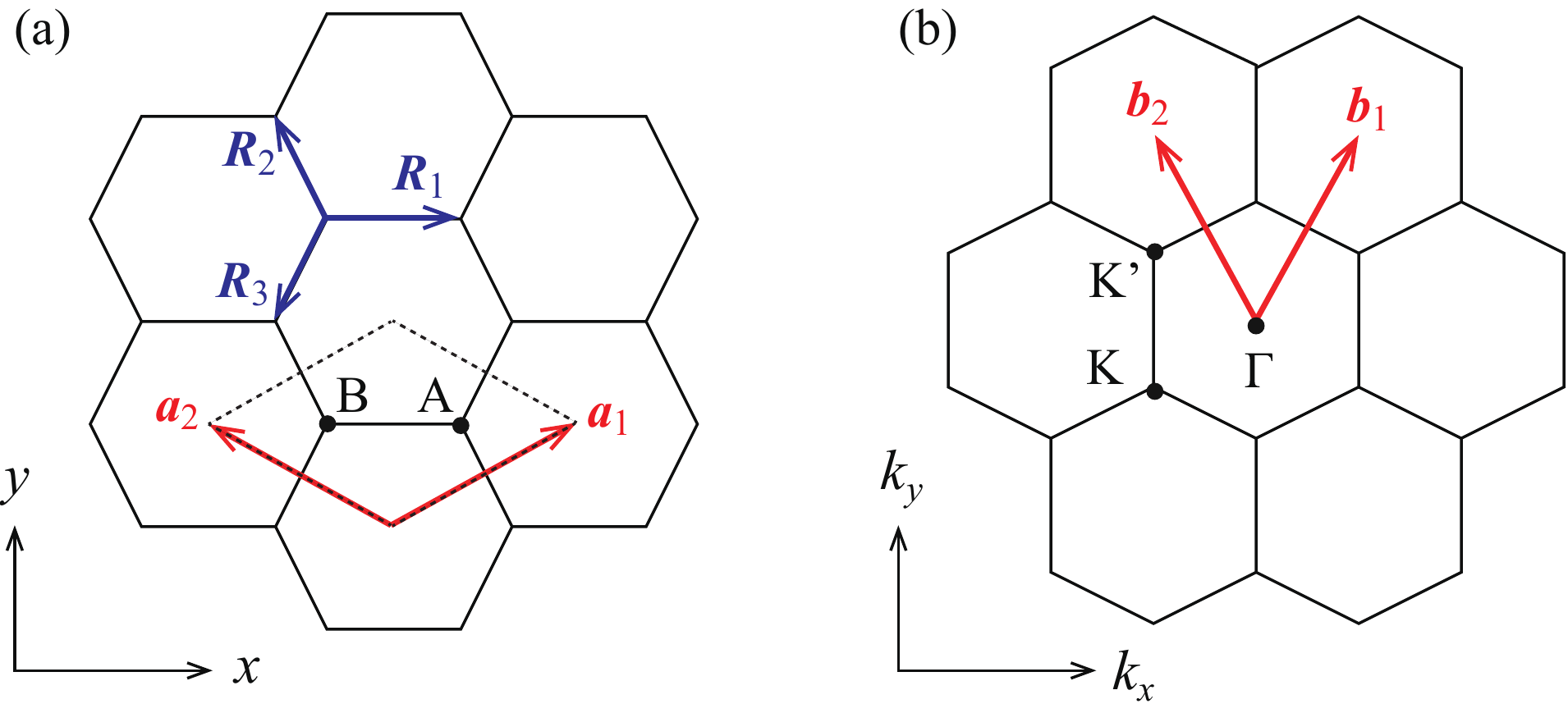}
  \end{center}
\caption{(a) Lattice structure of graphene and two primitive translation vectors 
given by ${\bm a}_1$ and ${\bm a}_2$. The unit cell is described by a hexagon 
containing two carbon atoms, A and B. ${\bm R}_j$ ($j=1,2,3$) are three vectors directed from 
a B atom to the three nearest neighbor A atoms. (b) Reciprocal lattice of graphene. 
${\bm b}_1$ and ${\bm b}_2$ are the reciprocal lattice vectors. The center of the hexagon (the first Brillouin zone) 
is called the $\Gamma$ point and the vertices of the hexagon are called K and K' points.}
\label{fig:A01}
\end{figure}

Figures~\ref{fig:A01}(a) and \ref{fig:A01}(b) show the real and reciprocal lattices of graphene, respectively. 
The two-dimensional principal lattice vectors are ${\bm a}_1=(-\sqrt{3}a/2, a/2)$ and ${\bm a}_2=(\sqrt{3}a/2, a/2)$
with $a\equiv|{\bm a}_1|=|{\bm a}_2|=0.246$nm. The unit cell contains two carbon atoms, A and B. 
The reciprocal lattice vectors are ${\bm b}=(-2\pi/\sqrt{3}a, 2\pi/a)$ and ${\bm b}_2=(2\pi/\sqrt{3}a, 2\pi/a)$, respectively.
In particular, the $\Gamma$, K, and K' points of the Brillouin zone are given by $\Gamma=(0,0)$, 
K$=(4\pi/\sqrt{3}a, 0)$, and K'$=(2\pi/\sqrt{3}a, 2\pi/a)$, respectively.

The $\pi$-orbital tight-binding Hamiltonian of graphene is given by
\begin{eqnarray}
H({\bm k})=\left(
\begin{matrix}
0 & H_{\rm AB}({\bm k}) \\
H_{\rm BA}(\bm k) & 0 \\
\end{matrix}
\right),
\label{eq:graphene_ham}
\end{eqnarray}
with the matrix elements $H_{AB}({\bm k})=\left(H_{BA}({\bm k})\right)^*$, which are given by
\begin{eqnarray}
H_{\rm AB}({\bm k})&\equiv& -\gamma_0\sum_{l=1}^3e^{i{\bm k}\cdot{\bm R}_l}\\
&=&-\gamma_0\left(
e^{ik_xa/\sqrt{3}}+2e^{-ik_xa/2\sqrt{3}}\cos\frac{k_ya}{2}
\right).
\label{eq:graphene_ham_elements}
\end{eqnarray}
Here, ${\bm R}_1=(a/\sqrt{3}, 0)$, ${\bm R}_2=(-a/2\sqrt{3}, a/2)$, and ${\bm R}_3=(-a/2\sqrt{3}, -a/2)$ are
the vectors going from a $B$ lattice point to three neighboring A lattice points, as shown in Fig.~\ref{fig:A01}.
$\gamma_0$ is the hopping integral between nearest-neighbor carbon atoms ($\pi$ orbitals, set to $\gamma_0=2.7$eV 
in the present study). By diagonalizing the Hamiltonian 
in Eq.~(\ref{eq:graphene_ham}), the energy dispersion relation of graphene can be obtained as
\begin{eqnarray}
E_{\pm}({\bm k})=\pm\gamma_0\sqrt{1+4\cos\frac{\sqrt{3}ak_x}{2}\cos\frac{ak_y}{2}+4\cos^2\frac{ak_y}{2}}
\label{eq:graphene_ene}
\end{eqnarray}
with the gapless points (so-called Dirac points) at the K and K' points. In Eq.~(\ref{eq:graphene_ene}), the sign
$\pm$ represents the conduction ($+$) and valence ($-$) bands, respectively. As is well known, graphene has
the linear dispersion relation $E_{\pm}({\bm k})\approx\pm\hbar v|{\bm k}|$ with $v=\sqrt{3}a\gamma_0/2\hbar$ around the K and K' points.

Here, zigzag-edged SWCNTs (z-SWCNTs) are considered as an example of pure 1D semiconductors.
The Hamiltonian of a z-SWCNT can be obtained by imposing the periodic boundary condition along the $y$ axis to the graphene as follows.
\begin{eqnarray}
H_q(k)=\left(
\begin{matrix}
0 & H_{\rm AB}^{(q)}(k) \\
H_{\rm BA}^{(q)}(k) & 0 \\
\end{matrix}
\right)
\label{eq:graphene_ham}
\end{eqnarray}
with the matrix elements $H_{\rm AB}^{(q)}(k)=\left(H_{\rm BA}^{(q)}(k)\right)^*$,
\begin{eqnarray}
H_{\rm AB}^{(q)}(k)
=-\gamma_0e^{ika/\sqrt{3}}\left(
1+2e^{-i\sqrt{3}ak/2}\cos\frac{\pi q}{n}
\right),
\label{eq:zCNT_ham_elements}
\end{eqnarray}
where $k$ is the wavenumber along the tube-axial direction, 
$q=0,1,\cdots,2n-1$ is the discrete wavenumber along the circumferential direction, and
$n$ is a natural number ($n=1, 2, \cdots, \infty$) specifying the unique structure of a particular z-SWCNT.
Herein, a z-SWCNT with index $n$ is represented as ($n$, 0) CNT in 
accordance with customary practice.

The energy dispersion relations $E_{\pm}^{(q)}(k)$ of 
the conduction $(+)$ and valence $(-)$ bands can be expressed as~\cite{rf:hamada, rf:saito}
\begin{eqnarray}
E_{\pm}^{(q)}(k)=\pm \gamma_0\sqrt{1+4\cos\left(\frac{ka_z}{2}\right)\cos\left(\frac{q\pi}{n}\right)+4\cos^2\left(\frac{q\pi}{n}\right)},\\
\left(q=0,1,\cdots, 2n-1 \quad{\rm and}~ -\pi/a_z<k<\pi/a_z\right)\nonumber.
\label{eq:TB_dispersion}
\end{eqnarray}
Here, $a_z=0.426$ nm is the unit cell length of the z-CNTs. 
A ($n$, 0) CNT includes $4n$ carbon atoms in the unit cell and its diameter $d_{\rm t}$ is given by 
$d_{\rm t}=\frac{na_z}{\sqrt{3}\pi}$.

z-SWCNTs can be either metallic or semiconducting depending on whether or not $n$ is a multiple of 3, respectively.
For the case of metallic z-SWCNTs ($n$~mod~$3=0$), two pairs of lowest-conduction (LC) and highest-valence (HV) bands 
$E_{\pm}^{(q)}(k)$ are specified by the following two values of $q$, respectively.
\begin{eqnarray}
q= \left\{
    \begin{array}{c}
      q_1\equiv 2n/3\\
      q_2\equiv 4n/3
    \end{array}
  \right. \quad {\rm for}\quad n~{\rm mod}~3=0.
\label{eq:pc0}
\end{eqnarray}
For the case of semiconducting z-SWCNTs ($n$~mod~$3\neq{0}$), the two pairs of LC and HV bands are respectively 
specified by 
\begin{eqnarray}
q= \left\{
    \begin{array}{c}
      q_1\equiv (2n+1)/3\\
      q_2\equiv (4n-1)/3
    \end{array}
  \right. \quad {\rm for}\quad n~{\rm mod}~3=1
\label{eq:pc1}
\end{eqnarray}
and
\begin{eqnarray}
q= \left\{
    \begin{array}{c}
      q_1\equiv (2n-1)/3\\
      q_2\equiv (4n+1)/3
    \end{array}
  \right. \quad {\rm for}\quad n~{\rm mod}~3=2.
\label{eq:pc2}
\end{eqnarray}
As can be seen in Eqs.~(\ref{eq:pc0})-(\ref{eq:pc2}), both the LC and HV bands will have two-fold degeneracy ($q_1$ and $q_2$) 
for a given $n$. 

In the long-wavelength limit ($k\to{0}$), the Hamiltonian matrix of z-SWCNT in Eq.~(\ref{eq:zCNT_ham_elements}) can be 
approximately expressed as
\begin{eqnarray}
H_{\rm AB}^{(q)}(k)\approx
e^{i\theta_k}\left\{\left(\Delta_q+\frac{\hbar^2k^2}{2m_q}\right)+i\hbar{v_q}k\right\}
\label{eq:k0_ham}
\end{eqnarray}
with $\theta_k=ka/\sqrt{3}$ and
\begin{eqnarray}
\Delta_q&=&-\gamma_0\left\{1+2\cos\frac{\pi q}{n}\right\}\\
m_q^{-1}&=&\frac{a_z^2\gamma_0}{2\hbar^2}\cos\frac{\pi q}{n}\\
v_q&=&\frac{a_z\gamma_0}{\hbar}\cos\frac{\pi q}{n}.
\end{eqnarray}

\begin{figure}[t]
  \begin{center}
  \includegraphics[keepaspectratio=true,width=70mm]{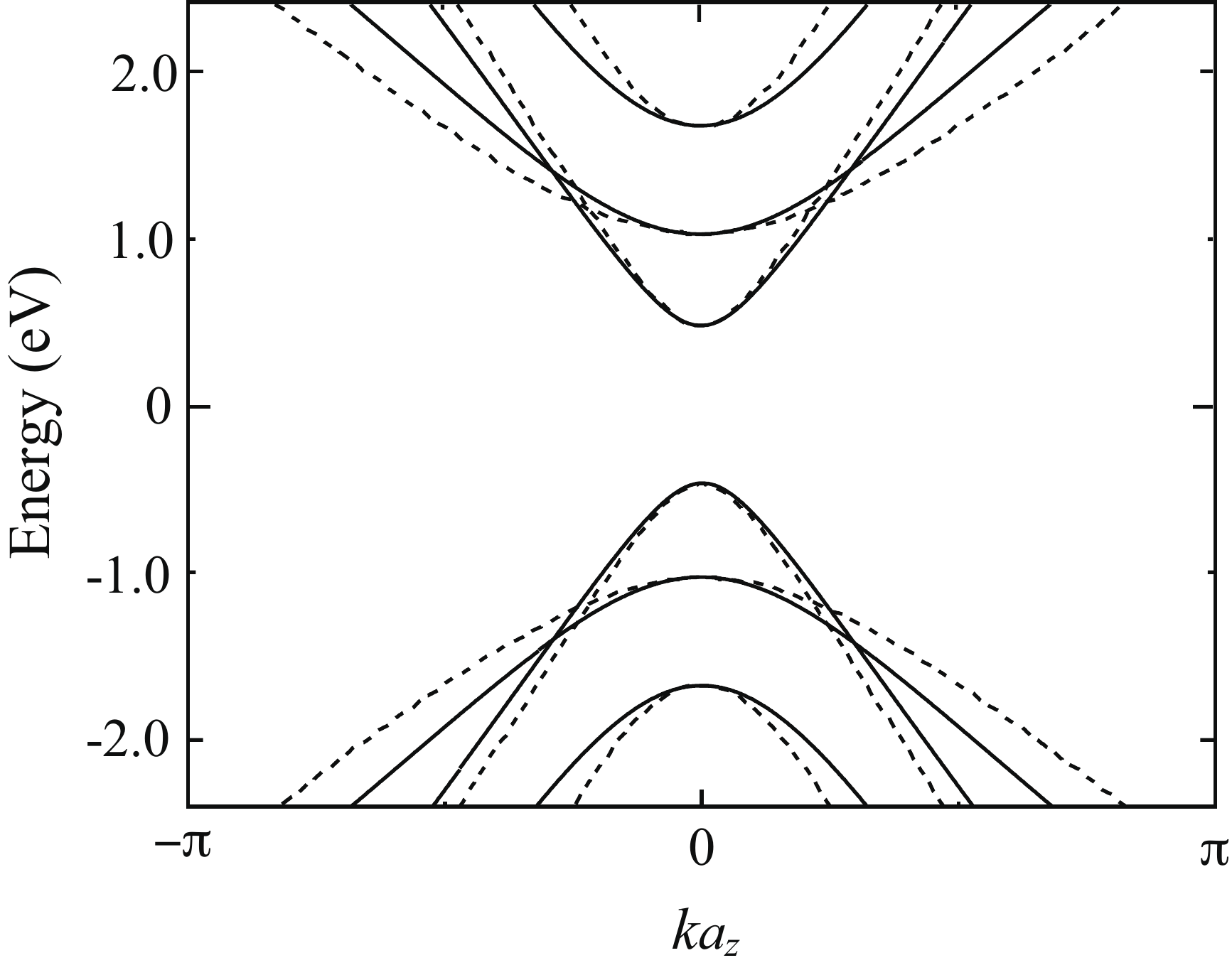}
  \end{center}
\caption{Energy bands around the Fermi energy ($\epsilon_{\rm F}=0$~eV) of a (10,0) z-SWCNT calculated using
the $\pi$-orbital tight-binding model (solid curves) and the 1D Dirac electron model (dashed curves).
Each curve is two-fold degenerate. 
The lowest conduction bands and the highest valence bands are denoted as 
the discrete wavenumber  $(q_1, q_2)=(7,10)$ along the circumferential direction of the (10,0) z-SWCNT, respectively. Similarly, 
the 2nd-lowest conduction bands and the 2nd-highest valence bands are denoted $(q_1, q_2)=(6,14)$, and the 3rd-lowest conduction
bands and the 3rd-highest valence bands are denoted as $(q_1, q_2)=(8,12)$.}
\label{fig:A02}
\end{figure}

The unitary transform of the Hamiltonian in Eq.~(\ref{eq:k0_ham}) can now be performed as
\begin{eqnarray}
{\tilde H}_q(k)&\equiv&U_k^\dagger H_q(k)U_k\\
&=&\left(
\begin{matrix}
\Delta_q+\frac{\hbar^2k^2}{2m_q} & \hbar{v_q}k \\
\hbar{v_q}k  & -\Delta_q-\frac{\hbar^2k^2}{2m_q} \\
\end{matrix}
\right)
\label{eq:1d_ham}
\end{eqnarray}
using the unitary matrix
\begin{eqnarray}
U_k=\frac{1}{\sqrt{2}}\left(
\begin{matrix}
1 & i \\
e^{-i\theta_k} & -ie^{-i\theta_k}  \\
\end{matrix}
\right).
\label{eq:k0_ham}
\end{eqnarray}

For the case of $\hbar|k|\ll 2\sqrt{m_q\Delta_q}$ (small wavenumber) and $\Delta_q/m_qv_q^2\ll 1$ (small band gap), 
the effective Hamiltonian in Eq.~(\ref{eq:1d_ham}) can be approximately described by
\begin{eqnarray}
{\tilde H}_q(k)\approx\left(
\begin{matrix}
\Delta_q & \hbar{v}_qk \\
\hbar{v}_qk  & -\Delta_q \\
\end{matrix}
\right).
\label{eq:1d_dirac_ham}
\end{eqnarray}
By diagonalizing the Hamiltonian ${\tilde H}_q(k)$ in Eq.~(\ref{eq:1d_dirac_ham}), the energy dispersion relations 
$E_{k}^{(\pm)}$ of the conduction $(+)$ and valence $(-)$ bands of 1D Dirac electrons can be obtained as
\begin{eqnarray}
E_{\pm}^{(q)}(k)=\pm\sqrt{\Delta_q^2+(\hbar v_qk)^2}.
\label{eq:1d_dirac_ene}
\end{eqnarray}

As an example, Fig.~\ref{fig:A02} shows the energy bands $E_{\pm}^{(q)}$ of the (10,0) z-SWCNT calculated using
the $\pi$-orbital tight-binding model (solid curves) and the 1D Dirac electron model (dashed curves).
The dashed curves are in excellent agreement with the solid curve in the vicinity of $k=0$.
For a (10, 0) z-SWCNT, the energy difference $\Delta E$ between the bottom of the LC band and the 
second-lowest conduction band is $\Delta E=0.557$~eV. Herein, the focus is on the low-energy excitation regime, 
in which the thermal energy $k_{\rm B}T$ is much lower than $\Delta E$. In the low-energy excitation regime,
the thermoelectric properties of z-SWCNTs can be explained in terms of single-band 1D Dirac electrons in
the LC and HV bands denoted as $q=q_1, q_2$ in Eqs.~(\ref{eq:pc1}) and (\ref{eq:pc2}). Here, it is assumed that 
the effects of possible mixing between the two LC bands due to impurity scattering can be ignored. This assumption
is valid under the condition that the characteristic momentum contributing to the formation of the bound state due to 
impurity potential is much smaller than the momentum difference between the two LC (HV) bands with $q=q_1$ and
$q_2$, as discussed in our previous work~\cite{rf:TY-HF}. 
Thus, in the main text of the present study, the focus is only on the LC and HV bands and the subscript or subscript 
$q=q_1$ and $q_2$ is dropped from the 1D Dirac Hamiltonian $h_0(k)$ in Eq.~(\ref{eq:1d_dirac_ham_k}).

\section{Derivation of Eq.~(\ref{eq:j_Q_x})~\label{sec:B}}
This appendix derives Eq.~(\ref{eq:j_Q_x}). By substituting Eq.~(\ref{eq:A}) into Eq.~(\ref{eq:dAdt}), the energy current $J_{\rm Q}$ can be rewritten as
\begin{eqnarray}
J_{\rm Q}&=&\int_{-\infty}^{\infty}\!\!\! dx\left\{ \frac{\partial\Psi^\dagger(x,t)}{\partial t}\left(\frac{\sin Qx}{2Q}H(x)\right)\Psi(x,t)\right.\nonumber\\
& &\left.+\Psi^\dagger(x,t)\left(\frac{\sin Qx}{2Q}H(x)\right)\frac{\partial\Psi(x,t)}{\partial t}\right\}+{\rm H.c.}
\label{eq:j_Q_01}
\end{eqnarray}
Applying Eqs.~(\ref{eq:sch_eq1}) and (\ref{eq:sch_eq2}) to Eq.~(\ref{eq:j_Q_01}), $J_{\rm Q}$ becomes
\begin{eqnarray}
J_{\rm Q}&=&\frac{i}{\hbar}\int_{-\infty}^{\infty}\!\!\! dx\frac{\sin Qx}{2Q}
\left\{(H(x)\Psi(x,t))^\dagger (H(x)\Psi(x,t))\right.\nonumber\\
& &\left.-(\Psi^\dagger(x,t)H(x))(H(x)\Psi(x,t))\right\}
+{\rm H.c.}
\label{eq:j_Q_02}
\end{eqnarray}
As the 1D Dirac Hamiltonian with a disorder potential $U(x)$ is given by $H(x)=-i\hbar\sigma_x\frac{\partial}{\partial x}+\Delta\sigma_z+U(x)$ 
in Eq.~(\ref{eq:Hx}), Eq.~(\ref{eq:j_Q_02}) can be rewritten as
\begin{eqnarray}
J_{\rm Q}&=&v\int_{-\infty}^{\infty}\!\!\! dx\frac{\sin Qx}{2Q}
\left\{\frac{\partial\Psi^\dagger(x,t)}{\partial x}\sigma_x H(x)\Psi(x,t)\right.\nonumber\\
& &\left.+\Psi^\dagger(x,t)\sigma_x\frac{\partial(H(x)\Psi(x,t))}{\partial x}\right\}
+{\rm H.c.}\nonumber\\
&=&v\int_{-\infty}^{\infty}\!\!\! dx\frac{\cos Qx}{2}\Psi^\dagger(x,t)\sigma_xH(x)\Psi(x,t)+{\rm H.c.}
\label{eq:j_Q_03}
\end{eqnarray}
In the limit of $Q\to 0$, Eq.~(\ref{eq:j_Q_x}) in the main text can be obtained as
\begin{eqnarray}
J_{\rm Q}=\frac{v}{2}\int_{-\infty}^{\infty}\!\!\! dx\Psi^\dagger(x,t)(\sigma_xH(x)+H(x)\sigma_x)\Psi(x,t).
\label{eq:j_Q_04}
\end{eqnarray}
Recently, Ogata and Fukuyama gave a general expression of Eq.~(\ref{eq:j_Q_x}) for 
multi-band disorder systems based on the L{\" u}ttinger-Kohn representation~\cite{rf:ogata-fukuyama2018}.

\section{Absence of vertex correction~\label{sec:C}}

\begin{figure}[t]
  \begin{center}
  \includegraphics[keepaspectratio=true,width=50mm]{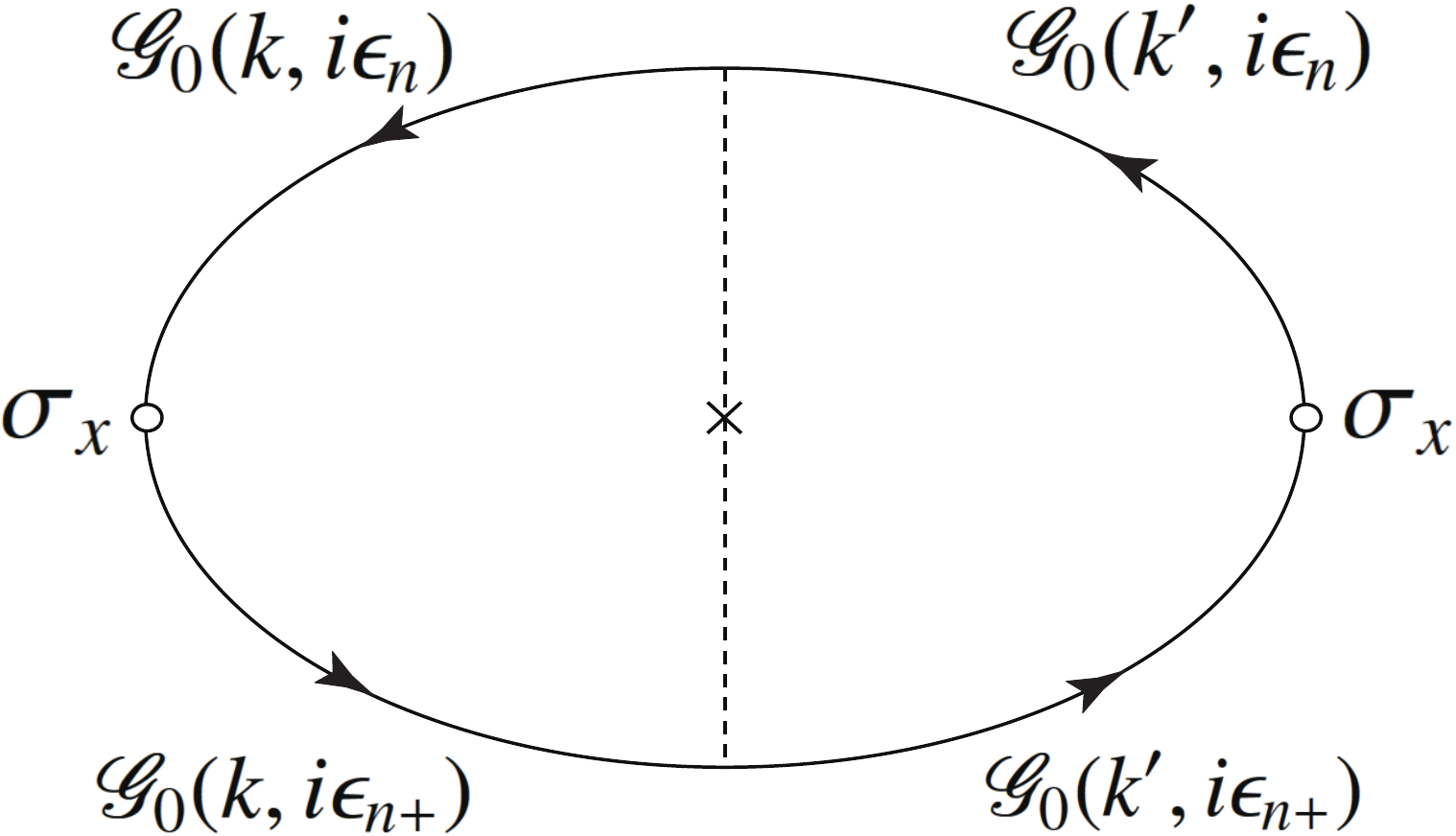}
  \end{center}
\caption{Diagram for the correlation function $\chi_{12}(i\omega_\lambda)$ in Eq.~(\ref{eq:chi_12}) with the lowest-order 
vertex correction. The dashed line with a cross represents an impurity potential $U$, the solid curves denote the unperturbed
thermal Green's functions ${\mathscr G}_0$ of 1D Dirac electrons, and the circles are the $x$-component of the Pauli matrix 
$\sigma_x$.}
\label{fig:vertex}
\end{figure}

In this appendix, a vertex correction for the current operator.. Up to the lowest order of $U$, $\sigma'_x(k)$ in Eq.~(\ref{eq:sigma'x})
is expressed as
\begin{eqnarray}
\sigma'_x(k)=\left\langle
U(k-k'){\mathscr G}_0(k,i\epsilon_{n+})\sigma_x{\mathscr G}_0(k',i\epsilon_n)U(k'-k)
\right\rangle_{\rm imp}
\end{eqnarray}
where we used ${\mathscr G}(k,k',i\epsilon_n)\approx\delta_{k,k'}{\mathscr G}_0(k,i\epsilon_n)$.
We now assume that the impurity potential is a short range and diagonal, that is, $U(q)$ is given by
\begin{eqnarray}
U(q)=\frac{U_0}{N}\sum_{\langle{l}\rangle}e^{-iqx_l}
\end{eqnarray}
with $U_0=(u_0I+u_z\sigma_z)$, where $I$ is the 2$\times$2 identity matrix. In this case, $\sigma'_x(k)$ is rewritten as
\begin{eqnarray}
\sigma'_x(k)=\sum_k\left(
U_0{\mathscr G}_0(k,i\epsilon_{n+})\sigma_x{\mathscr G}_0(k',i\epsilon_n)U_0\right)
f(k-k')
\end{eqnarray}
where $f(q)\equiv\left\langle \left(\frac{1}{N}\sum_{\langle{j}\rangle}e^{iqx_j}\right)\left(\frac{1}{N}\sum_{\langle{l}\rangle}e^{-iqx_l}\right)\right\rangle_{\rm imp}$.
Thus, the vertex correction $\Gamma$ for the correlation function in Eq.~(\ref{eq:chi_12}) is given by
\begin{eqnarray}
\Gamma&\propto&\sum_{k,k'}{\mathscr G}_0(k,i\epsilon_{n+})\sigma_x{\mathscr G}_0(k,i\epsilon_n)U{\mathscr G}_0(k',i\epsilon_{n+})\sigma_x{\mathscr G}_0(k',i\epsilon_n)U\nonumber\\
&=&\sum_k\frac{(A(k)\sigma_x-i\omega_\lambda\Delta\sigma_y)(u_0I+u_z\sigma_z)}{D(k,i\epsilon_{n+})D(k,i\epsilon_n)}\nonumber\\
& &\times\sum_{k'}\frac{(A(k')\sigma_x-i\omega_\lambda\Delta\sigma_y)(u_0I+u_z\sigma_z)}{D(k',i\epsilon_{n+})D(k',i\epsilon_n)}
\label{eq:vertex_corr}
\end{eqnarray}
where $A(k)\equiv i\epsilon_{n+}i\epsilon_n+\Delta^2+(\hbar vk)^2$. The scattering process in Eq.~(\ref{eq:vertex_corr}) is described in Fig.~\ref{fig:vertex}.

For a 1D Dirac electron, the unperturbed thermal Green's function ${\mathscr G}_0(k,i\epsilon_n)$ of is given by
\begin{eqnarray}
{\mathscr G}_0(k,i\epsilon_n)=\frac{i\epsilon_n+\hbar vk\sigma_x+\Delta\sigma_z}{D(k,i\epsilon_n)}.
\end{eqnarray}
with $D(k,i\epsilon_n)\equiv (i\epsilon_n)^2-(\Delta^2+(\hbar vk)^2)$ and then the numerator in Eq.~(\ref{eq:vertex_corr}) can be calculated as
\begin{eqnarray}
& &(A(k)\sigma_x-i\omega_\lambda\Delta\sigma_y)(u_0I+u_z\sigma_z)\nonumber\\
& &\times (A(k')\sigma_x-i\omega_\lambda\Delta\sigma_y)(u_0I+u_z\sigma_z)\nonumber\\
&=&\left[(u_0A(k)+u_z\omega_\lambda\Delta)\sigma_x-i(u_0\omega_\lambda\Delta+u_zA(k))\sigma_y\right]\nonumber\\
& &\times\left[(u_0A(k')+u_z\omega_\lambda\Delta)\sigma_x-i(u_0\omega_\lambda\Delta+u_zA(k))\sigma_y\right]\nonumber\\
&\approx&(u_0^2-u_z^2)\left[A(k)A(k')-i\omega_\lambda\Delta(A(k)-A(k')) \right].
\end{eqnarray}
Thus, the vertex correction in Eq.~(\ref{eq:vertex_corr}) vanishes for $u_z=\pm u_0$, corresponding to 
the disorder potential in Eq.~(\ref{eq:U_example}), {\it i.e.}, the two cases of $u_z=u_0$ and $u_z=-u_z$ correspond to
$U_1$ with $u_{11}=2u_0$ and $U_2$ with $u_{22}=2u_0$, respectively.

\section{Spectral conductivity near the mobility edges~\label{sec:D}}
As shown in Eq.~(\ref{eq:spectral_conductivity_ana}), the spectral conductivity $\alpha(E)$ is expressed as
\begin{eqnarray}
\alpha(E)\propto \frac{1}{{\rm Im}(\kappa_1\kappa_2)}{\rm Re}\left\{
\frac{2\kappa_1\kappa_2+\kappa_1^*\kappa_2+\kappa_1\kappa_2^*}{\sqrt{\kappa_1\kappa_2}}
\right\}
\label{eq:alpha_E2}
\end{eqnarray} 
within the SCBA. In this appendix, it is proven that the spectral conductivity behaves as $\alpha(E)\propto(E-E_{\rm c})$ near the mobility edge $E=E_{\rm c}$.

In the case of $E\lessapprox \Delta$, $\kappa_1$ and $\kappa_2$ are given by
\begin{eqnarray}
\kappa_1&=&-\kappa_1^{\rm R}+i\kappa_1^{\rm I},\\
\kappa_2&=&\kappa_2^{\rm R}+i\kappa_2^{\rm I}
\end{eqnarray} 
where $\kappa_1^{\rm R}>0$ and $\kappa_2^{\rm R}>0$. Using these expression, we obtain the following relations.
\begin{eqnarray}
{\rm Im}(\kappa_1\kappa_2)=\kappa_1^{\rm I}\kappa_2^{\rm R}-\kappa_1^{\rm R}\kappa_2^{\rm I},
\label{eq:Imk1k2}
\end{eqnarray} 
\begin{eqnarray}
\kappa_1^*\kappa_2+\kappa_1\kappa_2^*=2(-\kappa_1^{\rm R}\kappa_2^{\rm R}+\kappa_1^{\rm I}\kappa_2^{\rm I}),
\label{eq:k1k2+k1k2}
\end{eqnarray} 
and
\begin{eqnarray}
\sqrt{\kappa_1\kappa_2}=i\sqrt{\kappa_1^{\rm R}\kappa_2^{\rm R}\left(1+z\right)}
\label{eq:sqrtk1k2}
\end{eqnarray} 
with
\begin{eqnarray}
z\equiv \frac{\kappa_1^{\rm I}\kappa_2^{\rm I}}{\kappa_1^{\rm R}\kappa_2^{\rm R}}
-i\frac{\kappa_1^{\rm I}\kappa_2^{\rm R}-\kappa_1^{\rm R}\kappa_2^{\rm I}}{\kappa_1^{\rm R}\kappa_2^{\rm R}}.
\end{eqnarray} 
Because of $\kappa_1^{\rm I}\ll\kappa_1^{\rm R}$ and $\kappa_2^{\rm I}\ll\kappa_2^{\rm R}$ ({\it i.e.}, $z\ll 1$) neat the mobility edge
$E=E_{\rm c}$, Eq.~(\ref{eq:sqrtk1k2}) near $E=E_{\rm c}$ can be approximately expressed as
\begin{eqnarray}
\sqrt{\kappa_1\kappa_2}\approx i\sqrt{\kappa_1^{\rm R}\kappa_2^{\rm R}}\left(
1+\frac{z}{2}-\frac{z^2}{8}
\right)
\end{eqnarray} 
and eventually
\begin{eqnarray}
\frac{1}{\sqrt{\kappa_1\kappa_2}}\approx -i\frac{1}{\sqrt{\kappa_1^{\rm R}\kappa_2^{\rm R}}}
\left(
1-\frac{z}{2}+\frac{3z^2}{8}
\right).
\label{eq:1sqrtk1k2}
\end{eqnarray}

Substituting Eqs.~(\ref{eq:Imk1k2}), (\ref{eq:k1k2+k1k2}) and (\ref{eq:1sqrtk1k2}) into Eq.~(\ref{eq:alpha_E2}),
we can straightforwardly obtain
\begin{eqnarray}
\alpha(E)\propto\frac{2}{(\kappa_1^{\rm R}\kappa_2^{\rm R})^{3/2}}\kappa_1^{\rm I}\kappa_2^{\rm I}
\end{eqnarray} 
near the mobility edge $E=E_{\rm c}$. As can be seen in Fig.~\ref{fig:08}(a) and can be analytically verified from Eq.~(\ref{eq:e_eq_1_gen}) 
and Fig.~\ref{fig:07} as Im~$\sigma_1\propto -\sqrt{E-E_{\rm c}}$ together with 
$\sigma_1\sigma_2=\eta_1\eta_2$, the DOS is proportional to $\sqrt{E-E_{\rm c}}$ in the vicinity of
$E=E_{\rm c}$, and $\kappa_1^{\rm I}$ and $\kappa_2^{\rm I}$ near $E=E_{\rm c}$ show the behaviors 
$\kappa_1^{\rm I}\propto\kappa_2^{\rm I}\propto \sqrt{E-E_{\rm c}}$. Therefore, $\alpha(E)$ behaves as
\begin{eqnarray}
\alpha(E)\propto(E-E_{\rm c})
\end{eqnarray} 
near the mobility edge $E=E_{\rm c}$.

\end{document}